%% file: draft.tex
\documentclass[usenatbib]{mn2e}
\usepackage{graphicx, bm, amssymb, aas_macros, natbib}
\usepackage[colorlinks=true, citecolor=blue, urlcolor=blue,
linkcolor=blue]{hyperref}
\usepackage{color}
\input{extra_preamble.tex}

\title[The Local Group at high $z$]
{
The Local Group as a time machine:
studying the high-redshift Universe with nearby galaxies
} \author[M. Boylan-Kolchin et al.]  {Michael
  Boylan-Kolchin$^1$\thanks{$\!$mbk@astro.as.utexas.edu}, Daniel
  R. Weisz$^{2}$\thanks{$\!$Hubble Fellow}, Benjamin D. Johnson$^3$, \newauthor
  James S. Bullock$^4$, Charlie Conroy$^3$, Alex Fitts$^1$\\
\noindent $\!\!^1$Department of Astronomy and Joint Space-Science Institute,
University of Maryland, College Park, MD 20742-2421, USA\\
\noindent $\!\!^2$Astronomy Department, Box 351580, University of Washington,
Seattle, WA, USA\\
\noindent $\!\!^3$Harvard-Smithsonian Center for Astrophysics, 60 Garden St.,
Cambridge MA 02138, USA \\
\noindent $\!\!^4$Department of Physics and Astronomy, University of California
at Irvine, Irvine, CA 92697, USA}
\begin{document}

 \pagerange{\pageref{firstpage}--\pageref{lastpage}} 
 \pubyear{2015}

\maketitle
\label{firstpage}
\begin{abstract}
  We infer the UV luminosities of Local Group galaxies at early cosmic times
  ($z \sim 2$ and $z \sim 7$) by combining stellar population synthesis modeling
  with star formation histories derived from deep color-magnitude diagrams
  constructed from \textit{Hubble Space Telescope} (\hst) observations.  Our
  analysis provides a basis for understanding high-$z$ galaxies -- including
  those that may be unobservable even with the \textit{James Webb Space
    Telescope} (\jwst) -- in the context of familiar, well-studied objects in
  the very low-$z$ Universe. We find that, at the epoch of reionization, all
  Local Group dwarfs were less luminous than the faintest galaxies detectable in
  deep \hst\ observations of blank fields. We predict that \jwst\ will observe
  $z \sim 7$ progenitors of galaxies similar to the Large Magellanic Cloud
  today; however, the \hst\ Frontier Fields initiative may already be observing
  such galaxies, highlighting the power of gravitational lensing. Consensus
  reionization models require an extrapolation of the observed blank-field
  luminosity function at $z \approx 7$ by at least two orders of magnitude in
  order to maintain reionization. This scenario requires the progenitors of the
  Fornax and Sagittarius dwarf spheroidal galaxies to be contributors to the
  ionizing background at $z \sim 7$.  Combined with numerical simulations, our
  results argue for a break in the UV luminosity function from a faint-end slope
  of $\alpha \sim -2$ at $\muv \la -13$ to $\alpha \sim -1.2$ at lower
  luminosities. Applied to photometric samples at lower redshifts, our analysis
  suggests that \hst\ observations in lensing fields at $z \sim 2$ are capable
  of probing galaxies with luminosities comparable to the expected progenitor of
  Fornax.
\end{abstract}


\section{Introduction} 

Faint galaxies below the detection limits of current observatories are necessary
contributors to the ionizing background in the galaxy-dominated models of cosmic
reionization that are currently favored. Deep-field observations with the
\textit{Hubble Space Telescope} (\hst) only reach $\approx 0.1\,L^*(z \sim 7)$,
while essentially all models of reionization require galaxies $10-1000$ times
fainter to contribute to the ionizing background \citep{alvarez2012,
  kuhlen2012a, duffy2014, robertson2015, bouwens2015a}. Probing these likely
drivers of reionization is a prime motivation of the \textit{James Webb Space
  Telescope} (\jwst), yet even \jwst\ is unlikely to reach the faintest (and
most numerous) of these sources (see Section~\ref{subsec:near-deep}).

Resolved-star observations of Local Group galaxies provide an alternate path for
learning about the faintest galaxies at the epoch of reionization, via
``archeological'' studies of their descendants. Local galaxies also provide
important benchmarks against which data at a variety of redshifts can be
compared.  Accordingly, observations in the Local Group have long informed our
understanding of faint galaxies at early times \citep{bullock2000, freeman2002,
  ricotti2005, madau2008, bovill2011, brown2012, benitez-llambay2015}. 

The relation between high-$z$ star formation rates (SFRs) and dark matter halo
masses for the progenitors of Local Group dwarfs is of particular interest for
understanding the reionization epoch. \citet[hereafter B14]{boylan-kolchin2014}
showed that the requisite inefficient star formation in low-mass dark matter
halos at $z=0$ proves difficult to reconcile with expected physics of the
high-redshift Universe in general and reionization models in particular, as
these often require relatively efficient star formation in low-mass halos at
early times. Indeed, \citet{madau2014b} suggested that the reionization-era
progenitors of present-day dwarf galaxies may have been among the most efficient
sites of conversion of gas into stars on galactic scales. Efficient galaxy
formation at high redshift has low-redshift implications, however: based on
counts of remnants of atomic cooling halos from $z=8$ that survive to $z=0$ in
simulated Local Groups, B14 argued that the UV luminosity function (LF) should
break to a shallower value than the observed $\alpha \approx -2$ for galaxies
fainter than $\muv \sim -14$ (corresponding to halo masses of
$\mvir \la 10^9\,\msun$).

\citet[hereafter W14]{weisz2014c} further demonstrated the power of combining
local observations with stellar population synthesis models in confronting LFs
in the high-redshift Universe: they showed that one can construct LFs that reach
several orders of magnitude fainter than is currently observable.  The results
of W14 indicate that UV LFs continue without a sharp truncation to \textit{much}
fainter galaxies [$\muv(z \sim 5) \approx -5$], showing how near-field
observations can inform our understanding of galaxies that are not directly
observable at present.

In this paper, we combine the approaches of B14 and W14. Using resolved star
formation histories (SFHs), we model the UV luminosities that Local Group
galaxies had at earlier epochs. We compare these to existing \hst\ data at
$z\sim 2$ (from UV dropout galaxies in blank fields and cluster lensing fields)
and at $z \sim 7$ (from the Ultra-Deep Field (HUDF) and the Frontier Fields). As
we show below, this approach holds the promise of both placing the faintest
observable galaxies at high redshift into a more familiar context and
understanding the possible role of well-studied Local Group galaxies in
high-redshift processes such as cosmic reionization.

\section{Modeling Local Group Galaxies at higher redshifts}
\label{sec:methods}
\subsection{Deriving $\muv(z)$}
\label{subsec:getting_muv}
\begin{figure*}
 \centering
 \includegraphics[scale=0.41, viewport=10 0 438 453]{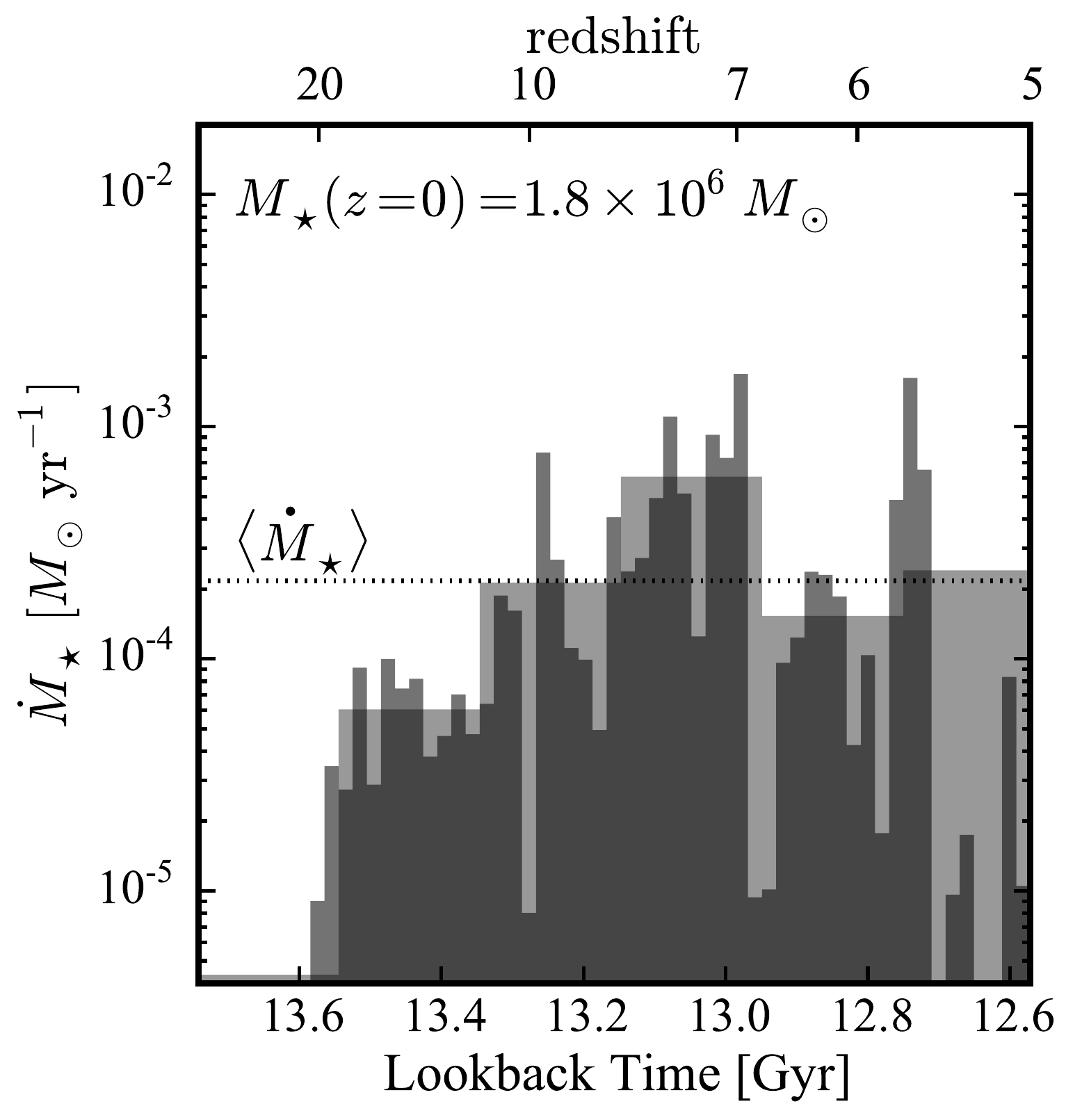}
 \includegraphics[scale=0.41, viewport=20 0 363 463]{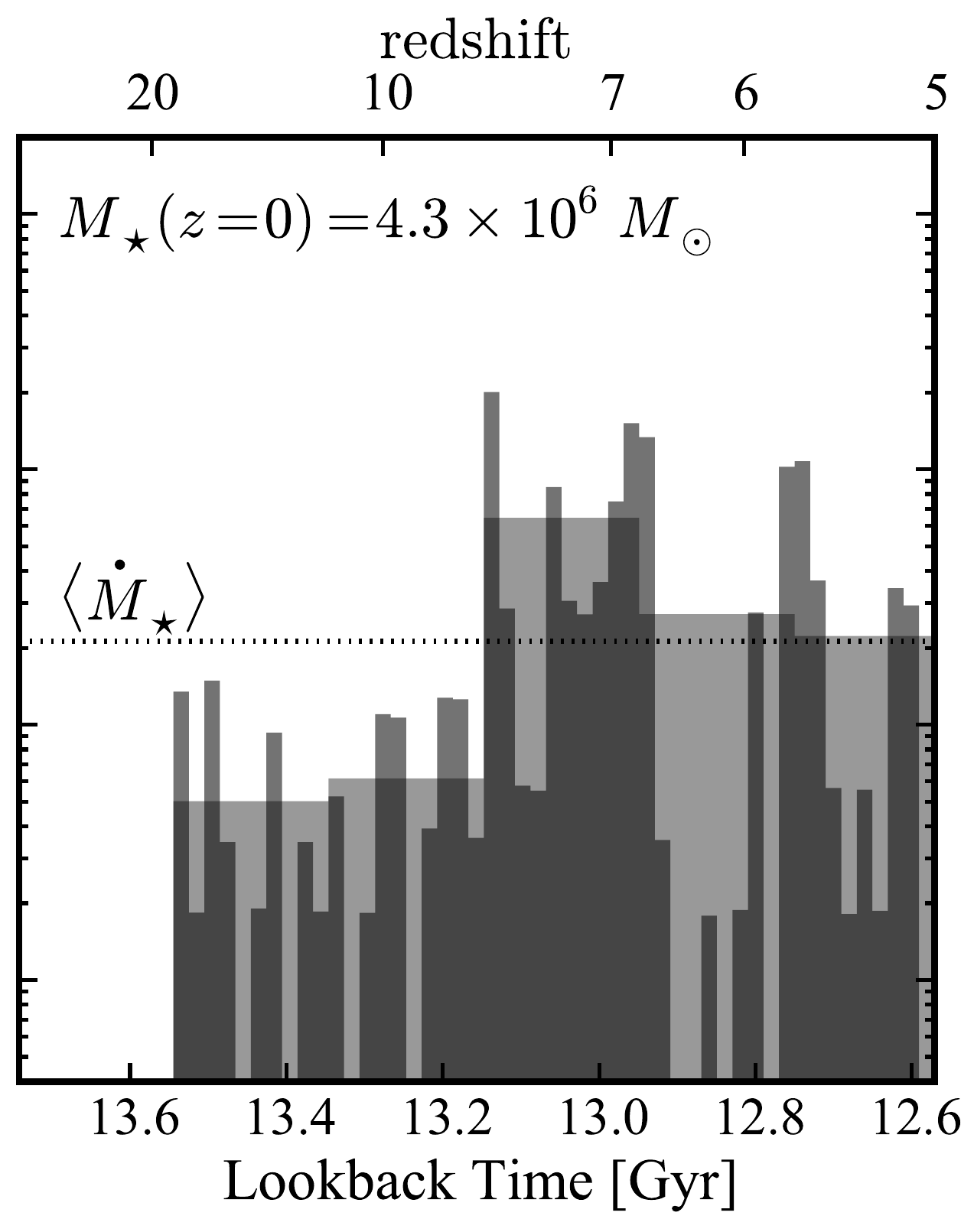}
 \includegraphics[scale=0.41, viewport=20 0 433 413]{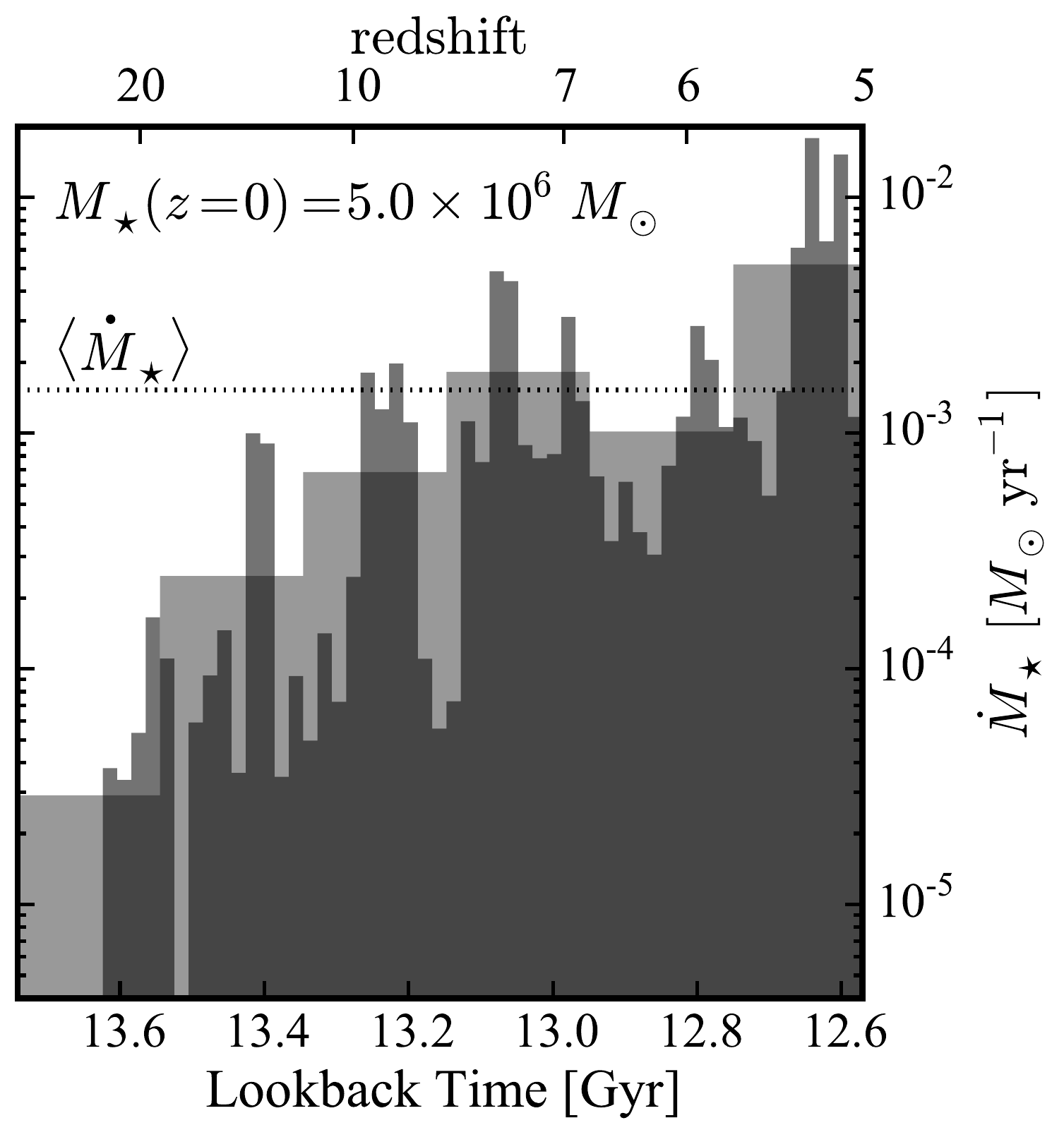}
 \caption{Star formation rates calculated in bins of 200 Myr (light gray
   histograms) and 20 Myr (dark gray histograms) for simulated galaxies from
   \citeauthor{onorbe2015} (2015; left) and Fitts et al. (in preparation; center
   and right) up to $z=5$. At the present day, each galaxy has
   $10^6 \la \mstar(z=0)/\msun \la 5\times 10^6$ and is hosted by a halo with
   $\mvir(z=0) =10^{10}\,\msun$. 
   The dotted horizontal line shows the mean SFR over the period plotted. 
   Averaged over 200 Myr periods, the SFRs appear to be
   increasing to $z=7$. On 20 Myr timescales, they are much burstier and
   fluctuate strongly. These simulated SFHs motivate the burst
   parametrizations we use in our modeling. 
 \label{fig:sfr}
}
\end{figure*}

To interpret our observations of Local Group galaxies in a high-redshift
context, we follow the methodology described in W14. This procedure is
summarized here; for further details, see W14 and references therein. 

We begin with SFHs derived from resolved star color-magnitude diagrams (CMDs)
constructed from \hst\ imaging. The majority of the SFHs we use are based on
analysis of archival \hst/WFPC2 imaging; these data, the SFHs, and details on
how the SFHs were computed are presented in \citet{weisz2013} and
\citet{weisz2014a}. We select Local Group galaxies that span a wide range in
present-day stellar mass ($2\times 10^5-2 \times 10^9\,\msun$) and have
\hst-based CMDs that extend below the oldest main sequence turn-off (MSTO),
which enables a precise constraint on the stellar mass at all epochs back to
$z=5$. The archival WFPC2 data meet the MSTO criteria for galaxies within
$\sim 400$ kpc of the MW, and of these, we select a representative subset. We
supplement this dataset with SFHs of Leo A (presented in \citealt{cole2007}) and
IC 1613 (presented in \citealt{skillman2014}), which are based on newer imaging
that includes the ancient MSTO. Our sample includes SFHs for all known bright
($L_V>10^{5}\,L_{V,\odot}$) MW satellite galaxies with the exception of Sextans
(for which there is no sufficiently deep \hst\ imaging), as well as several
Local Group dwarf irregular galaxies. SFHs from Local Group galaxies are
``non-parametric'' in the sense that they are the sum of simple stellar
populations (e.g., \citealt{dolphin2002}) as opposed to imposed analytic
prescriptions (e.g., $\tau$ models).

The SFH of each galaxy is input into the Flexible Stellar Population Synthesis
code \citep{conroy2009a, conroy2010a} in order to generate UV and $V$-band flux
profiles over time. In both of these steps, we use the Padova stellar evolution models
\citep{girardi2002, girardi2010} and a \citet{kroupa2001} initial mass function
(IMF). We assume a constant metallicity of $0.2\,Z_{\odot}$ and no internal dust
extinction (see the discussion in Section~\ref{subsec:uncertainties}). The
latter assumption appears to be in broad agreement with the apparent dust-free
spectral energy distributions of faint, high-redshift galaxies (e.g.,
\citealt{bouwens2012, dunlop2013}) and with expectations from simulations
\citep{salvaterra2011}. Finally, in order to generate the simulated fluxes as a
function of redshift, we require the simulated $V$-band luminosity at $z=0$ to
match observations of Local Group galaxies as listed in
\citet{mcconnachie2012}. This normalization accounts for the fact that the \hst\
field of view does not always cover the entire spatial extent of a nearby
galaxy.

We initially generate predicted fluxes assuming the fiducial SFH, i.e., a
constant SFH over each time bin. However, as suggested by a variety of
simulations\footnote{$\!$Not all simulations agree on this point: for example,
  \citet{vogelsberger2014b} were able to match many observable properties of
  dwarf galaxies with galaxy formation models that result in much smoother star
  formation rates as a function of time.} \citep{stinson2007, ricotti2008,
  governato2012, zolotov2012, teyssier2013, dominguez2014, power2014,
  onorbe2015} and observations (e.g., \citealt{van-der-wel2011, kauffmann2014}),
the SFHs of low-mass galaxies at high redshift are likely to fluctuate on timescales
($\sim$ 10-100 Myr) that are shorter than what is directly available from the fossil
record (which provides a time resolution of $\sim$ 10-15\% of a given lookback
time for CMDs that include the oldest MSTO).
 
To account for short duration bursts of star formation, we have modified the
fiducial SFHs of our Local Group sample to include short duration ($\sim$ 10-100
Myr) bursts. That is, in each time bin, we insert bursts with a specified
amplitude and duration that are stochastically spaced in time, with the
requirement that the total stellar mass formed in each time bin matches that of
the fiducial SFH. For the purpose of this exercise, we have adopted two
representative burst models: 200 Myr duration with amplification factor of 5 and
20 Myr duration with amplification factor of 20. In both cases, we assume that
80\% of the star formation occurs in the bursting phase. Each modified SFH is
therefore a series of 20 (200) Myr bursts with an amplitude of 20 (5) times the
average SFH instead of the fiducial constant SFH in a given time bin (as noted
above, each time bin has a width of $\sim$ 10-15\% of the corresponding lookback
time).

The choice of these parameters is guided by cosmological
simulations. Figure~\ref{fig:sfr} shows examples of such episodic SFHs for three
simulated dwarf galaxies at $z>5$. Each was run using {\tt Gizmo}
\citep{hopkins2015} with meshless finite-mass hydrodynamics at very high force
and mass resolution and using the {\tt FIRE} implementations of galaxy formation
and stellar feedback \citep{hopkins2014}. The dwarf in the left panel is a
version of the ``dwarf-early'' galaxy presented in \citet{onorbe2015} and
\citet{wheeler2015}, while the dwarfs in the center and right panels are of
halos with nearly identical masses selected from a $35\,\mpc$ volume and
resimulated at very high resolution (from A. Fitts et al., in preparation). The
average SFRs of these galaxies at $z>5$ are
$(0.1-1) \times 10^{-3}\,\msun\,{\rm yr}^{-1}$ and their stellar masses at $z=0$
are $(1.8-5)\times 10^{6}\,\msun$. 

Figure~\ref{fig:sfr} shows the early ($z \ge 5$) star formation in each halo,
with the light gray (dark gray) histogram showing the star formation averaged
over 200 (20) Myr periods. While the variation when averaged over 200 Myr
periods is typically a factor of 2--5, the variation on 20 Myr timescales is
frequently a factor of 10--20 and can even exceed 100. This is further
corroborated by SFRs for a variety of halos over the range $11 > z > 5$
presented in \citet{ma2015}. While the masses of their halos
[$\mvir(z=6) \sim 10^{10-10.5}\,\msun$] are generally larger than those
considered here [$\mvir(z=6) \sim 10^{8.5-9.5}$], the SFHs are qualitatively
similar in their very bursty, episodic nature. \\[0.3cm]


\subsection{From $\muv(z)$ to $P(\muv)$}
\label{subsec:pmuv}

The result of this modeling is a distribution of magnitudes as a function of
lookback time, $\muv(t)$, or redshift, $\muv(z)$, for each galaxy.  The
probability distribution $P(\muv)$ for each galaxy is then given by
$P(\muv) \propto \muv(z)\,P(z)$, i.e., it is a distribution over the modeled UV
flux as a function of time, weighted by the probability that the galaxy falls
into a sample with redshift selection function $P(z)$. The weights can
correspond to the photometric redshift distribution from an observational
sample, to some modification thereof, or to an arbitrary selection function.

In what follows, we adopt a Gaussian centered at $z=7$ with width
$\sigma_z=0.475$ as the redshift selection function $P(z)$ for our $z \sim 7$
galaxies. This produces a good match to the photometric redshift distribution
from \citet{finkelstein2014} (after excluding the secondary peak at $z \sim 1.4$
originating from detections of a 4000 ${\rm \AA}$ break rather than the Lyman
break). Similarly, we use a Gaussian centered at $z=1.91$ with a width of
$\sigma_z=0.21$ as the redshift selection function $P(z)$ for our $z \sim 2$
sample, as this reproduces the photometric redshift distribution from
\citet{oesch2010a}. At both $z\sim 7$ and $z \sim 2$, the width of $P(z)$ is
smaller than the time resolution of the observed SFHs.

\begin{figure*}
 \centering
 \includegraphics[scale=0.485]{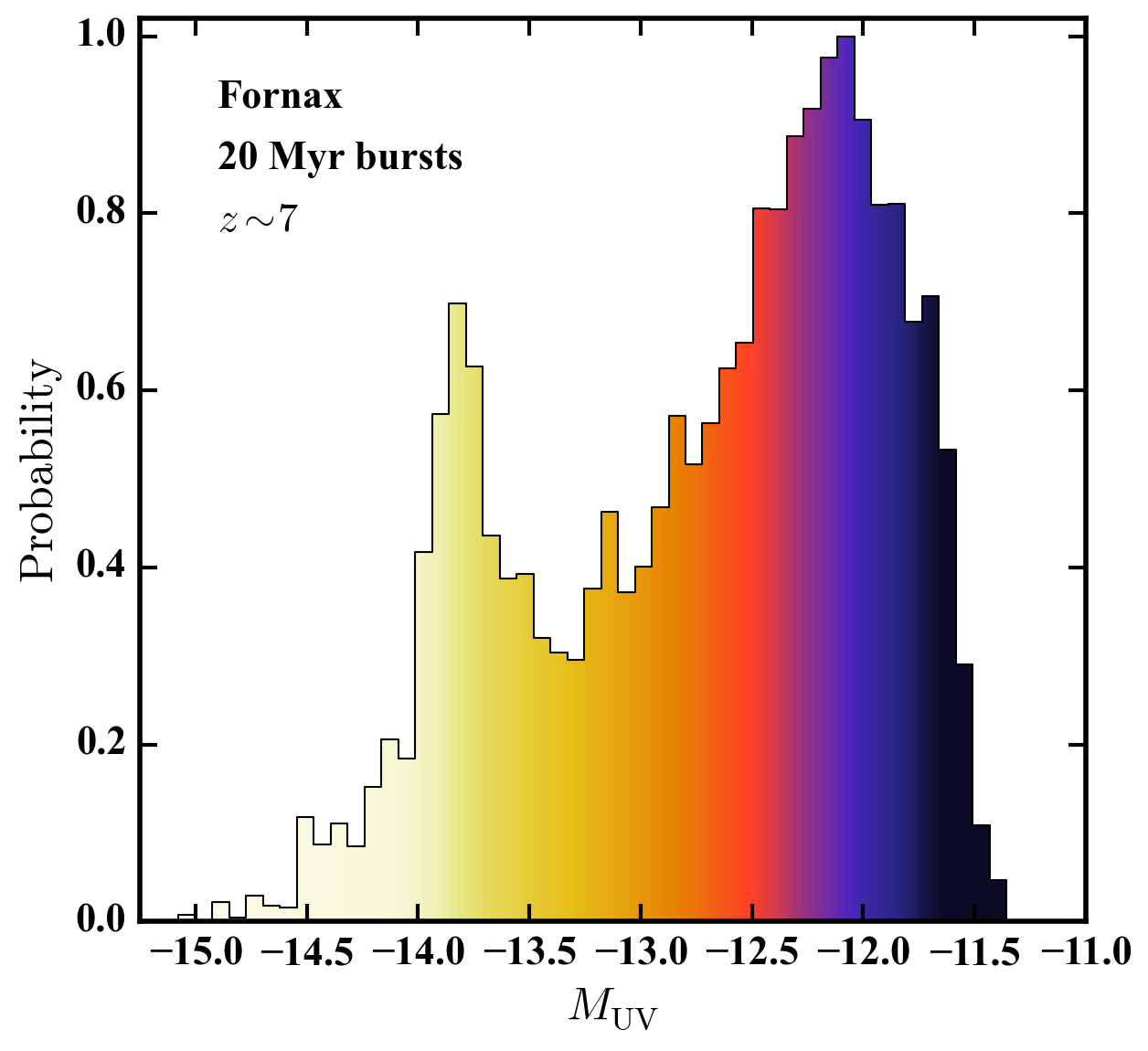}
 $\;\;\;\;\;\;\;$
 \includegraphics[scale=0.485]{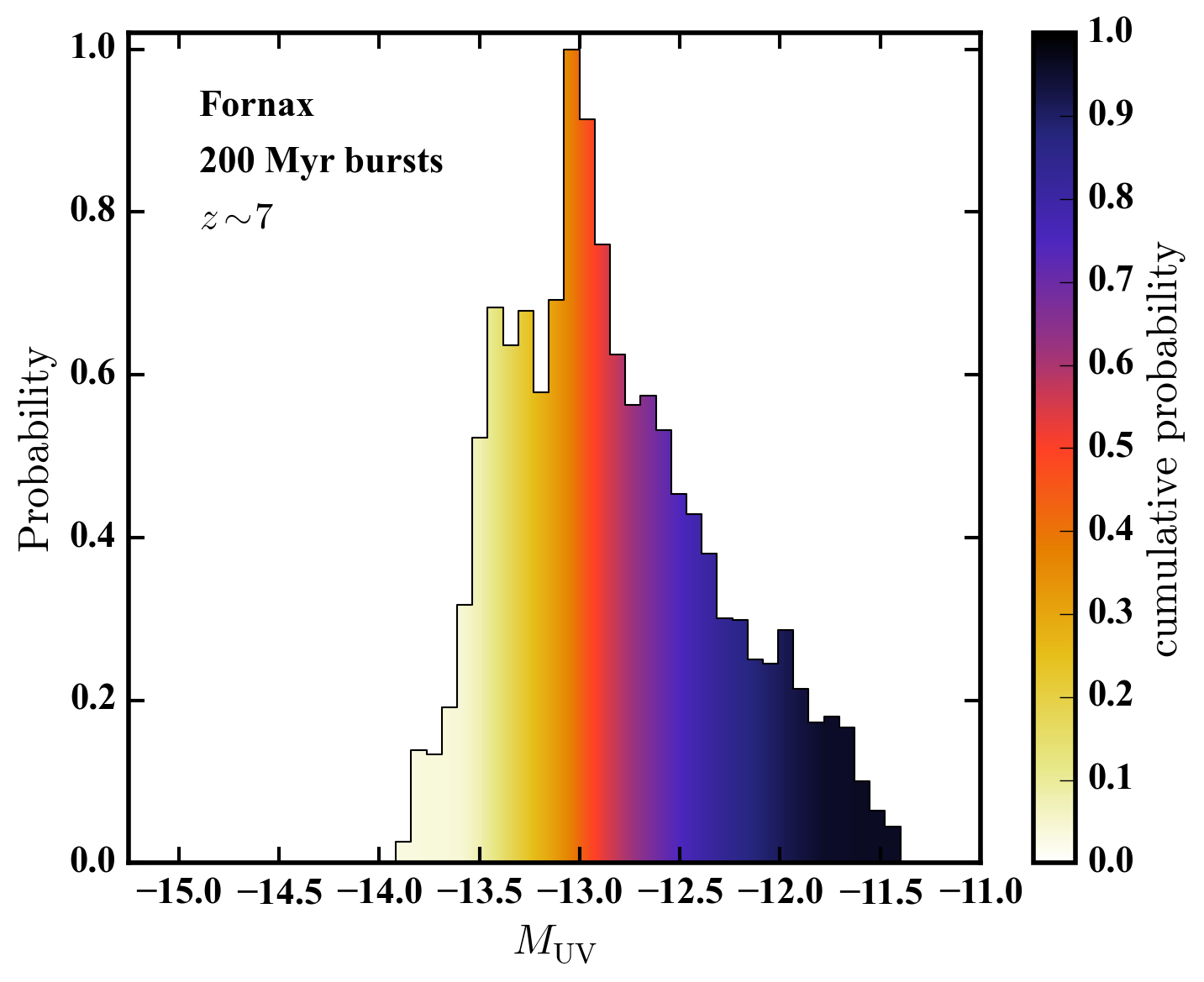}
 \caption{Probability distribution for $\muv$ of Fornax, given the photometric
   redshift distribution of Finkelstein et al. and assuming that Fornax's
   high-redshift star formation occurred in bursts with a characteristic
   timescale of 20 Myr (left) or 200 Myr (right). The colors indicate the
   cumulative probability distribution in each panel. 
   In the case of 20 Myr bursts, the frequent and relatively high-amplitude
   bursts result in a bimodal probability distribution function (with peaks corresponding
   to the burst and inter-burst periods). The 200 Myr bursts, which are modeled
   with smaller burst amplitudes, result in a distribution that is less broad
   and is unimodal. This is a direct consequence of our smaller assumed burst
   amplitude for 200 Myr bursts; the resulting SFH is therefore much closer to
   the fiducial constant SFH. Taking the instantaneous value of $\muv$ at $z=7$
   in each realization results in a cumulative distribution that is very similar
   to that shown here in each case. 
 \label{fig:fornax_burst_muv}
}
\end{figure*}

Figure~\ref{fig:fornax_burst_muv} shows the probability distribution $P(\muv)$
for the Fornax dwarf spheroidal (dSph) galaxy, computed from 1000 realizations
of its SFH, at $z \sim 7$. The left panel shows the case of 20 Myr bursts
while the right panel shows the result for 200 Myr bursts. The distributions
are shown as shaded histograms; the shading gives the value of the cumulative
distribution function $P(<\!\muv)$ for each $\muv$. Taking the instantaneous value
of $\muv$ at $z=7$ in each realization results in a cumulative distribution that
is very similar to that shown here in each case.

While the full range of $\muv$ spanned by the distributions is set by the burst
fluctuation amplitude and is therefore roughly the same in the case of 20 or 200
Myr bursts, the distributions themselves differ substantially. 20 Myr bursts
result in a bimodal distribution with a prominent peak $\sim 0.5$ magnitudes
from the faint end of the distribution and a secondary peak $\sim 1.0$
magnitudes from the bright end of the distribution. The median value falls at
$\sim 1$ magnitude from the faint end of the distribution. In the case of 200
Myr bursts, the distribution is noticeably different: there is a much stronger
peak near the median of the distribution, with relatively high probabilities of
being up to half a magnitude brighter or a magnitude fainter than the median
value of $\muv$. The net result is that the median probability in the case of
200 Myr bursts is approximately 0.5 magnitudes brighter than in the (likely more
realistic) case of 20 Myr bursts (e.g., \citealt{weisz2012a}). This is similar
to the value of $\muv$ that would be obtained by assuming a constant SFR over
the entire period to $z=5$, i.e., the parameters we adopt for 200 Myr bursts act
mainly to distribute $\muv$ about the median value for the constant SFH.

\begin{table}
  \caption{
    Observed $V$-band magnitudes of
    Local Group dwarfs (column 2), along with modeled UV magnitudes at $z \sim
    2$ (column 3) and $z \sim 7$ (column 4). The $\muv$ values quoted in columns
    3 and 4 are the medians of the probability distributions, while the errors
    are the minimum range containing 68\% (95\%) of the
    cumulative probability for $\muv$. 
    \label{table:table1}
  }
\renewcommand{\arraystretch}{2}
  \begin{tabular*}{\columnwidth}{@{\extracolsep{\fill}} lrrr}
    \hline
    \hline
    Name & $M_{\rm V}(z=0)$ & $\muv(z \sim 2)$ & $\muv(z \sim 7)$ \\
    \hline
    Carina  &  $-9.1 \pm 0.5$  &  $-8.3^{+0.6\, (2.4)}_{-1.6\, (2.0)}$  
                                               &  $-8.9^{+0.9\, (1.3)}_{-0.9\, (1.5)}$ \\
    CVn I  &  $-8.6 \pm 0.2$  &  $-6.9^{+3.9\, (3.9)}_{-1.4\, (3.2)}$ 
                                               &  $-9.5^{+0.9\, (0.9)}_{-0.6\, (1.6)}$ \\
    Draco  &  $-8.8 \pm 0.3$ &  $-9.4^{+0.6\, (2.3)}_{-1.7\, (2.2)}$  
                                               &  $-9.2^{+0.9\, (0.9)}_{-0.5\, (1.4)}$ \\
    Fornax  &  $-13.4 \pm 0.3$  & $-12.1^{+0.9\, (0.9)}_{-0.6\, (1.6)}$  
                                               &  $-12.6^{+1.1\, (1.1)}_{-0.5\, (1.4)}$  \\
    IC 1613  &  $-15.2 \pm 0.2$  &  $-12.6^{+0.6\, (1.6)}_{-1.2\, (1.8)}$  
                                               &  $-13.4^{+0.9\, (1.0)}_{-0.7\, (1.6)}$  \\
    Leo A  &  $-12.1 \pm 0.2$  &  $-5.7^{+0.7\, (0.9)}_{-0.5\, (1.2)}$  
                                               &  $-10.5^{+1.0\, (1.0)}_{-0.6\, (1.3)}$ \\
    Leo I  &  $-12.0 \pm 0.3$  &  $-5.1^{+0.9\, (0.9)}_{-2.2\, (4.5)}$  
                                               & $-11.1^{+1.0\, (1.0)}_{-0.6\, (1.4)}$  \\
    Leo II  &  $-9.8 \pm 0.3$  &  $-8.7^{+1.2\, (1.5)}_{-0.8\, (2.2)}$
                                               &  $-8.8^{+0.7\, (0.7)}_{-0.5\, (1.5)}$  \\
    Leo T  &  $-8.0 \pm 0.5$  &  $-1.5^{+0.4\, (0.8)}_{-0.4\, (0.8)}$  
                                               &  $-7.9^{+0.9\, (0.9)}_{-0.6\, (1.5)}$  \\
    LMC  &  $-18.1 \pm 0.1$  &  $-15.6^{+0.8\, (0.9)}_{-0.6\, (1.6)}$  
                                               &  $-15.8^{+1.0\, (1.1)}_{-0.6\, (1.4)}$  \\
    Phoenix  &  $-9.9 \pm 0.4$  &  $-9.0^{+1.0\, (1.0)}_{-0.6\, (1.5)}$
                                               & $-10.0^{+0.8\, (0.8)}_{-0.6\, (1.5)}$ \\
    Sagittarius  &  $-13.5 \pm 0.3$  & $-5.9^{+0.4\, (0.8)}_{-0.4\, (0.8)}$  
                                               &  $-12.3^{+0.9\, (0.9)}_{-0.5\, (1.5)}$  \\
    Sculptor  &  $-11.1 \pm 0.5$  &  $-10.5^{+0.5\, (2.3)}_{-1.5\, (1.9)}$  
                                               &  $-11.2^{+0.6\, (1.3)}_{-1.2\, (1.3)}$  \\
    SMC  &  $-16.8 \pm 0.2$  &  $-14.3^{+0.8\, (1.3)}_{-1.0\, (1.8)}$  
                                               &  $-14.1^{+0.4\, (1.3)}_{-1.2\, (1.2)}$  \\
    Ursa Minor  &  $-8.8 \pm 0.5$  & $-6.2^{+1.7\, (1.6)}_{-2.9\, (5.1)}$ 
                                               &  $-8.9^{+1.0\, (1.0)}_{-0.6\, (1.5)}$  \\
    \hline
  \end{tabular*}
\end{table}

Given the expectations from observations and cosmological simulations, we adopt
20 Myr bursts as our default model for star formation and compute UV magnitudes
from the resulting probability distribution derived from 100 independent
realizations per galaxy. The results at $z\sim 2$ and $z\sim 7$ are listed in
Table~\ref{table:table1}. We quote median values and confidence intervals
comprising the minimum range in $\muv$ containing 68\% and 95\% of the
cumulative probability. Since virtually all of the galaxies have $P(\muv)$
distributions that look like Figure~\ref{fig:fornax_burst_muv}, these ranges are
usually asymmetric about the median.

\begin{figure}
 \centering
 \includegraphics[scale=0.56, viewport=30 0 400 420]{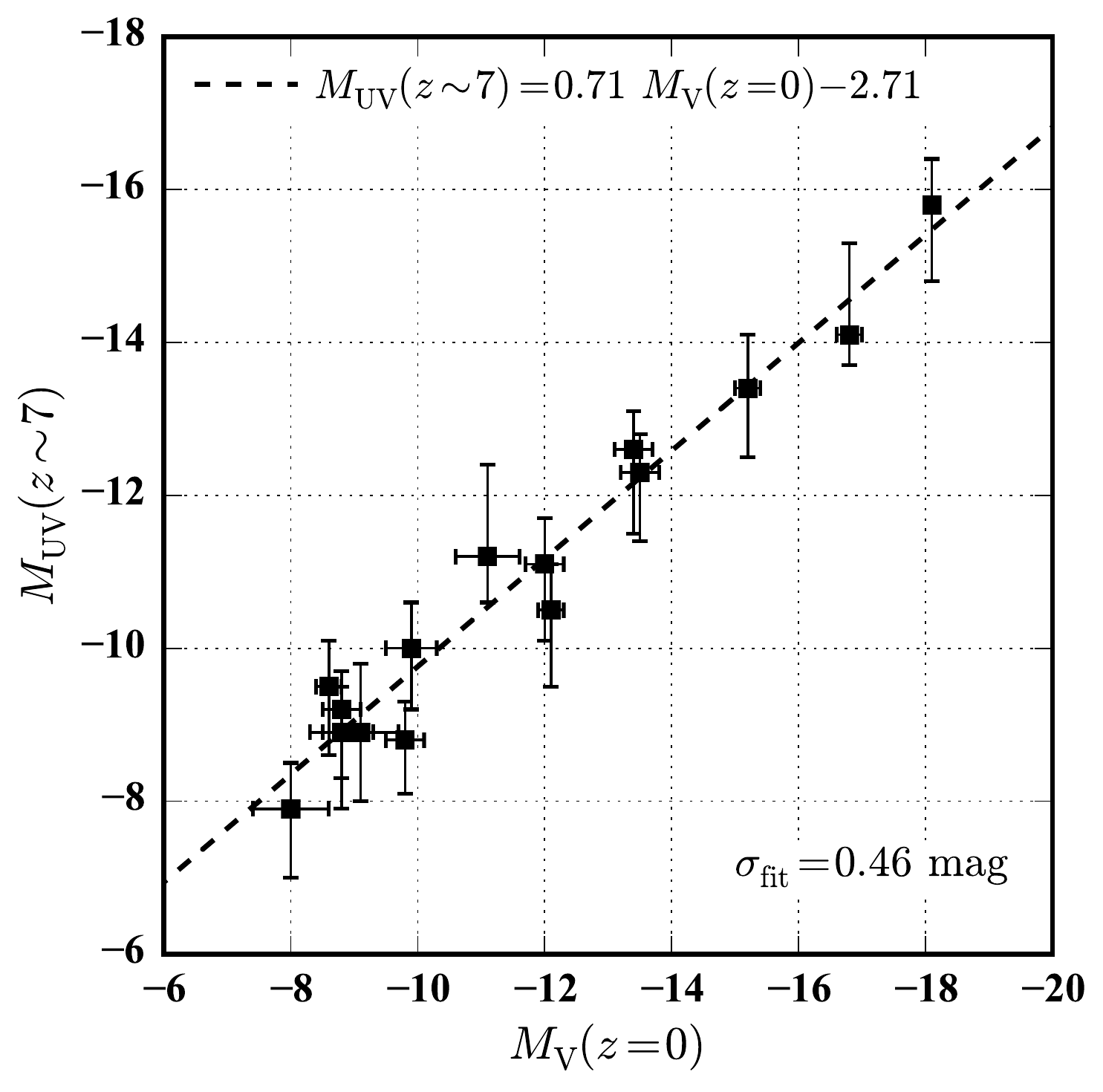}
 \caption{Observed $V$-band magnitude at $z=0$ versus modeled $z \sim 7$ UV
   magnitude. Error bars indicate minimum 68\% confidence intervals (see
   Sec.~\ref{subsec:pmuv} for more details). There is a well-defined, linear
   relation between the present-day $V$-band magnitude and the modeled
   high-redshift UV magnitude for the galaxies in our sample with a dispersion
   of $\sim 0.5$ magnitudes. The detailed form
   of this relation relies on the mode of star formation in galaxies. For the dwarf
   galaxies studied in this paper, we have assumed that star formation occurs
   predominantly in short-duration bursts. More massive galaxies are likely to
   maintain star formation over longer periods, which would result in a shallower
   relation between $\muv(z \sim 7)$ and $M_{\rm V}(z=0)$.
 \label{fig:uv_z0_z8}
}
\end{figure}

An immediate connection that we can make between high redshifts and current-day
quantities is to compare observed $V$-band magnitudes at $z=0$ with modeled
values of $\muv(z=7)$. Such a comparison is shown in
Figure~\ref{fig:uv_z0_z8}. Perhaps unsurprisingly, there is a clear correlation
between high-redshift UV luminosity and present-day $V$-band luminosity for the
dwarf galaxies studied here. The slope of this correlation is $0.71$, indicating
that the high-redshift UV luminosities span a narrower range than the $z=0$
$V$-band luminosities. While there is non-negligible scatter in this relation,
which is related to the diverse early SFHs of the low-mass dwarfs in particular
(e.g., figure 12 of \citealt{weisz2014a}), the correlation is well-defined over
the range of luminosities studied here. We caution against extrapolating this
relation to significantly higher luminosities, as it is unlikely to remain
linear indefinitely. The balance between $\muv(z \sim 7)$ and $M_V(z=0)$ is set
by the stellar mass formed at early times versus over a galaxy's lifetime, and
very massive galaxies today have predominantly ancient stellar populations
\citep{thomas2005}, which is more similar to low-mass classical dSphs in the MW
than to galaxies at intermediate mass (such as the MW itself).

For our sample of galaxies as a whole, $\muv(z=2)$ is reasonably well-correlated
with $\muv(z=7)$ and with $M_V(z=0)$. There are notable exceptions, however,
including Leo A, Leo I, Leo T, and Sagittarius. These galaxies show much lower
UV luminosities at $z\sim2$ than would be naively expected based on their
present-day luminosities. The origin of this difference lies in the SFHs of
these galaxies: while all of the galaxies in our sample have ancient
($z \sim 7$) star formation, the galaxies listed above appear to have gone through
relatively quiescent phases at lower redshifts (see figure~7 of
\citealt{weisz2014a}). These periods of quiescence result in dramatically
reduced UV luminosities at those times.

As discussed at the beginning of this section, we have only included a
representative set of Local Group galaxies.  They are mainly selected based on
diversity in stellar mass and morphology at $z=0$, and all have the requirement
of reaching below the oldest MSTO in order to provide the best possible
constraints on the stellar mass formed by $z=5$.  Future work analyzing the CMDs
of newer, deep \hst\ observations will enable us to include a more exhaustive
sample of Local Group galaxies.

\subsection{Uncertainties in, and limitations of, our approach}
\label{subsec:uncertainties}

There are a number of assumptions we must make in our modeling. For
completeness, we list several of them here.
\begin{itemize}
\item \textit{Star formation histories}: we have only used the best-fit SFH,
  i.e., we assume perfect knowledge of the stellar mass at $z=5$.  This is
  mitigated by the SFHs reaching the ancient MSTO, but it
  still can introduce uncertainties at the factor of $\la 2$ level into the
  total stellar mass formed.
\item \textit{Aperture corrections}: we assume the measured SFHs are
  representative of the entire galaxy. If the galaxies have strong population
  gradients, however, this assumption may be violated. The results of
  \citet{hidalgo2013} indicate that at least some Local Group dwarf irregular
  galaxies do have population gradients, with younger stellar populations being
  more centrally concentrated than older populations. Any bias in our
  calculations is therefore likely to be in the direction of underestimating the
  UV luminosities at early times, as most of the galaxies have observations that
  sample close to their centers.
\item \textit{Stellar IMF}: we assume a fully populated Kroupa IMF, but the IMF
  may not be Kroupa or fully populated. Furthermore, we are basing the SFH on
  the IMF of lower-mass stars, but it was the high-mass stars in the early
  Universe that produced the UV light \citep{zaritsky2012}.
\item \textit{Dust}: we do not make any corrections for dust in our modeling.
  This is unlikely to be a poor assumption for early-time progenitors of the
  Local Group dwarf galaxies we are considering, as the vast majority are very
  metal-poor even today. More massive, and vigorously star-forming, galaxies at
  the epoch of reionization may be dusty (see, e.g., \citealt{finkelstein2012,
    watson2015}). The galaxies studied here are expected to be low-metallicity
  and forming stars at rates of $10^{-4}-10^{-2}\,\msun\,{\rm yr^{-1}}$ at
  $z\sim7$, however, and are likely to be dust-poor \citep{fisher2014}.
  \item \textit{Metallicity}: we assume a constant metallicity of
    $0.2\,Z_{\odot}$ throughout our analysis. Variations around this choice are
    unlikely to affect our modeling of $\muv$ appreciably; see
    \citealt{johnson2013} for further details and a more expansive discussion of
    several of the uncertainties discussed here.
\item \textit{Distances}: Our present analysis ignores the difference in
  luminosity distance over the width of photometric redshift distribution
  $P(z)$. For our $z=7$ sample, the distance modulus only varies by $\sim 0.2$
  magnitudes from $z=6.5$ to $z=7.5$, meaning it will introduce at most a small
  correction to the effects we have modeled. The difference at $z \sim 2$ is
  $\sim 0.4$ magnitudes, which is still sub-dominant to the effects induced by
  bursty star formation.
\end{itemize}

One potentially important uncertainty is the merger histories of dwarf
galaxies. Our approach is inherently archeological in nature, as we are using
resolved SFHs of galaxies observed at $z=0$ to study the
high-redshift Universe. Our analysis yields modeled UV fluxes as a function of
time for the stars that are in the galaxy today, but those stars may have been
formed in multiple distinct progenitors and only assembled more recently. If,
for example, a typical merger history for our modeled galaxies were such that
the $z=0$ galaxy had $N$ equal-mass progenitors at $z=7$, then our inferred UV
luminosities at that epoch would be an overestimate by a factor of $N$. We
assess the likely distribution of progenitor masses in two ways, through the use
of abundance matching to dark-matter-only simulations (for a statistical
understanding) and through high-resolution hydrodynamic simulations (for
individual case studies).

Our first method relies on the {\tt ELVIS} suite of dark-matter-only simulations
of Local Group analogs \citep{garrison-kimmel2014}. We identify all objects
within 1.2 Mpc of either Local Group giant at $z=0$ (excluding the central subhalos,
which correspond to the MW and M31) and trace these objects back in time. Rather
than following only the most massive progenitor, we track \textit{all}
progenitors at each previous redshift. We find that, at $z\sim 7$, there is
typically a dominant progenitor in terms of mass: the next most massive
progenitor is more than a factor of two lower in mass than the most massive
progenitor at that epoch, on average. Based on our modeling, this indicates that
the main progenitor is likely to be at least one magnitude brighter in the UV
than any other progenitor. Since the SFHs are highly
episodic, this statement is true only in a time-averaged sense.

We can also address this issue directly through hydrodynamical
simulations. Specifically, we can use both an archeological and an instantaneous
approach to $z\sim 7$ star formation by asking (1) how many of the stars in the
galaxy at $z=0$ formed by $z=7$, and (2) what is the stellar content of the main
progenitor of the $z=0$ galaxy at $z=7$? The difference between these two
answers is a direct measure of the difference between resolved-star studies at
$z=0$ and observations at $z \sim 7$. We use the cosmological zoom-in
simulations described in Section~\ref{subsec:getting_muv} to perform this
test. In all cases, the main progenitor of the simulated dwarf contains
approximately 70\% or more of the stars in the $z=0$ galaxy that have formed by
$z = 7$.  The main progenitor is therefore already dominant by that time (see
also \citealt{dominguez2014}), and our archeological approach to the properties
of Local Group galaxy progenitors at high redshift should be a reasonable
approximation.

\section{High-redshift observations in a Local Group context}
\label{sec:highz}
With the results of Section~\ref{sec:methods}, we are now in position to
consider observations at high redshift in the context of well-observed galaxies
in the Local Group. Of particular interest is the nature of the faintest objects
observable in deep fields; in what follows, we discuss how the progenitors of
Local Group dwarf galaxies would appear at $z\sim 7$ and $z \sim 2$.

\subsection{Local Group galaxies at cosmic dawn}
\label{subsec:z7}

\begin{figure*}
 \centering
 \includegraphics[scale=0.57]{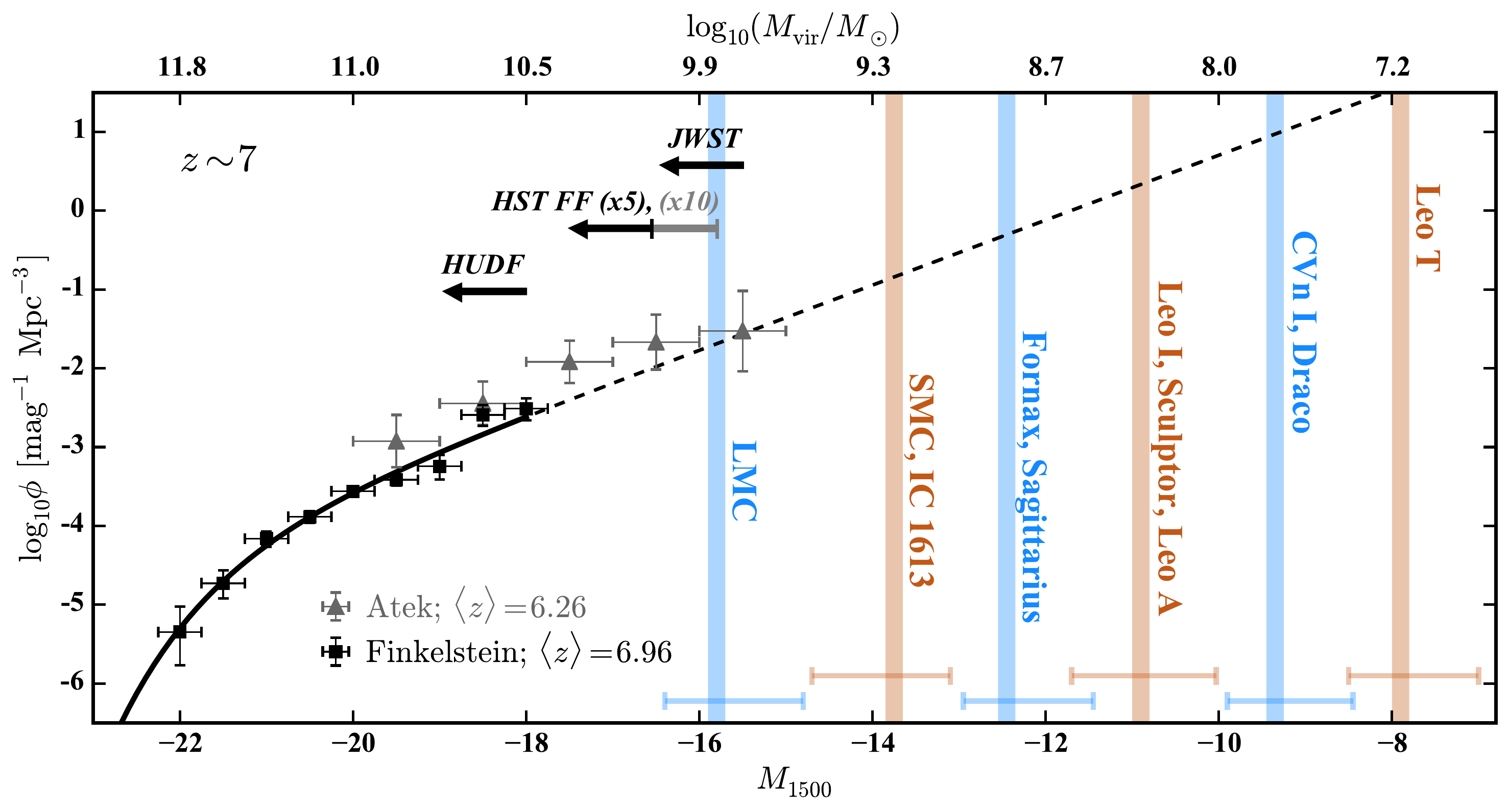}
 \caption{Local Group galaxies at $z\sim 7$. The data points show the observed
   UV luminosity function at $z \sim 7$ from Finkelstein et
   al. (\citeyear{finkelstein2014}; \hst, no lensing; black squares) and Atek et
   al. (\citeyear{atek2015}; \hst\ Frontier Fields; gray triangles). The
   best-fit \citet{schechter1976} function
   ($M^*=-21.03, \, \phi^*=1.57\times 10^{-4}\,{\rm mag^{-1}\,Mpc^{-3}},\,
   \alpha=-2.03$)
   from Finkelstein et al. is displayed as a solid curve, while its
   extrapolation to lower luminosities is shown as a dashed line. The
   observational limits are also shown for the HUDF (assuming a completeness
   limit of $\muv=-18$), the \hst\ \ff\ [assuming $m_{\rm lim}=28.7$ before
   lensing and a lensing magnification of 5 (1.75 magnitudes) or 10 (2.5
   magnitudes)], and \jwst\ (assuming $m_{\rm lim}=31.5$; this same depth would
   be reached in the Frontier Fields at $z=7$ with a magnification of 13.18 or
   2.8 magnitudes). The modeled $\langle \muv(z \sim 7) \rangle$ for various
   Local Group galaxies are plotted as vertical bands (with error bars giving
   $1\,\sigma$ uncertainties; see Table~\ref{table:table1}), indicating the
   power of local observations for interpreting deep-field data. Consensus
   reionization models require extrapolation to $\muv(z \sim 7) \approx -13$,
   corresponding to the brightest dSphs today (Fornax, Sagittarius).
  \label{fig:lf_plot_z7}
}
\end{figure*}
Figure~\ref{fig:lf_plot_z7} shows the observed UV LF of galaxies at $z \sim 7$
(from \citealt{finkelstein2014}; see also \citealt{mclure2013, schenker2013,
  schmidt2014, bouwens2015}). The approximate halo masses expected for a given
$\muv$ are shown on the upper horizontal axis; these are obtained by matching
cumulative number densities of galaxies and halos (assuming a \citet{sheth2001}
mass function for the latter). In the deepest blank field, the HUDF, \hst\ is
capable of reaching $\muv \approx -17.5$. Models of reionization in which
galaxies play the dominant role in maintaining an ionized intergalactic medium
require a major contribution from fainter galaxies, however, meaning an
important subset of the galaxies responsible for reionization have not yet been
directly observed. The luminosities of the faintest galaxies contributing to
reionization are unknown even theoretically, as this limit (often denoted
$\mlim$) depends on the slope of the UV LF and the escape
fraction of ionizing photons from these galaxies (among other factors). The
minimum $\muv$ necessary for maintaining reionization in models ranges from
$\mlim \approx -15$ to $\mlim \ga -10$ (e.g., \citealt{kuhlen2012a,
  robertson2013}).

Our calculations of $\muv (z \sim 7)$ for Local Group galaxies can therefore
provide crucial context for galaxies at cosmic dawn. We find that the
progenitor of the Large Magellanic Cloud (LMC) -- the brightest Local Group
dwarf galaxy in our sample both at $z=0$ and $z \sim 7$-- had
$\muv(z\sim 7) \sim -16$, beyond the capabilities of \hst\ in the HUDF\footnote{This
  prediction of $\muv(z\sim 7) \approx -16$ for the LMC puts it very close to
  the UV luminosity for the MW, as predicted by B14 through abundance matching,
  at that time. This is consistent with expectations from the {\tt ELVIS} suite
  (and other $N$-body simulations): the present-day most massive satellite is
  often only slightly lower (a factor of $1.5-2$) in mass than the main halo
  itself at $z=7$.}.  The Small Magellanic Cloud (SMC) and IC 1613 likely had
$\muv(z=7) \sim -13.8$; virtually all reionization models therefore require
progenitors of such galaxies to contribute to reionization. If
$\muv(z\sim 7) \approx -13$ galaxies are required to maintain reionization, as
is the case in many models, then the progenitors of the Sagittarius and Fornax
dSphs must contribute. And finally, if galaxies as faint as
$\muv(z \sim 7) \sim -9.5$ are required, then progenitors galaxies such as
Draco, CVn I, and Phoenix -- which have
$L_V(z=0) \sim (0.2-1)\times 10^{6}\,L_{V,\odot}$ -- may be necessary
contributors to reionization.

These results also provide context for what observations in the \hst\ \ff\ and
with \jwst\ will see. At $z \sim 7$, \ff\ can probe galaxies as faint as
$\muv \approx -16.5$, assuming magnifications of 5 in luminosity (1.75 mags);
such observations may approach the depth required for observing the main
progenitor of the LMC.  \jwst\ deep field observations are expected to have a
limiting magnitude of $m_{\rm AB}=31.5$ \citep{windhorst2006}, corresponding to
$\muv \sim -15.5$; \jwst\ is therefore likely to reveal the $z\sim 7$
progenitors of Magellanic irregulars. If a \ff-like campaign with \jwst\ could
obtain a factor of 10 in magnification (2.5 mags), it would observe objects as
faint as $\muv \approx -13$; this would just approach the sensitivity required
to observe the progenitors of Fornax and Sagittarius, potentially revealing the
faintest galaxies required for reionization. (We note that, based on the results
of W14, we do not expect a strong truncation in the LF at at $\muv \sim -13$ or
even at significantly fainter magnitudes; the idea of a limiting magnitude
$\mlim$ required for reionization is more of a mathematical construct than a
physical cut-off. However, this does not preclude the possibility that the LF
becomes shallower near $\mlim$; see Sec.~\ref{subsec:z7}.)  Progenitors of the
vast majority of Local Group dwarfs will remain unobservable even with a \jwst\
\ff-like project, however. This highlights the inherent difficulty of high-$z$
observations and the power of studying the high-$z$ Universe through its local
descendants.

\subsection{Near-field / deep-field connections}
\label{subsec:near-deep}
\begin{figure}
 \centering
 \includegraphics[scale=0.56, viewport=20 0 400 460]{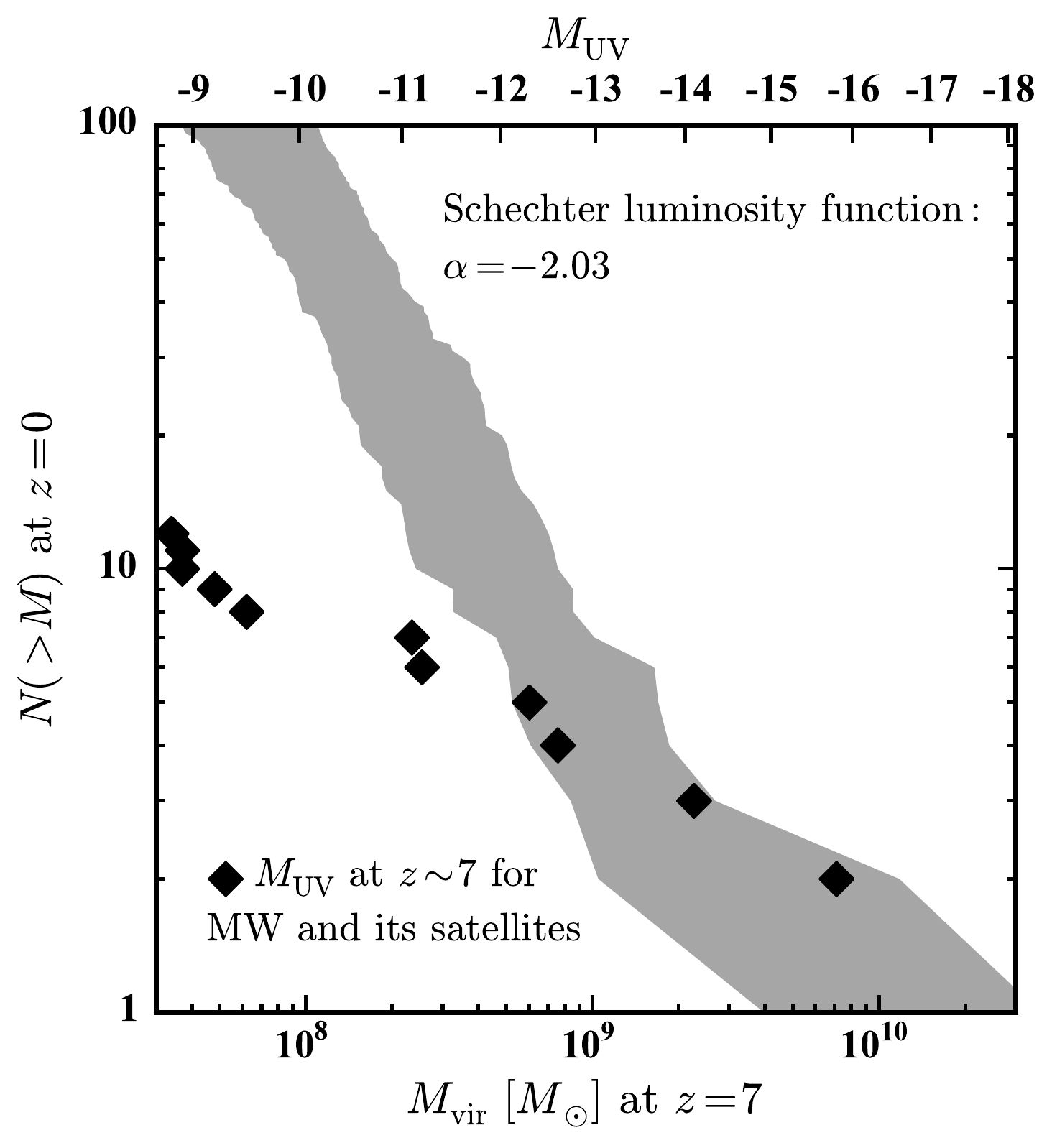}
 \caption{The $z=7$ mass function of the main progenitors of surviving $z=0$
   (sub)halos -- including the main progenitor of the MW itself -- within 300
   kpc of the Milky Way based on the {\tt ELVIS} simulations (shaded region). The
   upper horizontal axis gives the abundance-matched $\muv$; the data points
   show our modeled UV luminosity function for progenitors of the MW and its
   satellites at $z \sim 7$. The luminosity function from direct modeling of
   SFHs and from abundance matching are in good agreement for 
   $\muv \la -12.5$ ($\mvir \approx 7\times 10^{8}\,\msun$) but diverge for
   fainter galaxies (lower mass halos), perhaps indicating the need for a break
   in the UV luminosity function at $\muv \approx -13$ at $z \sim
   7$.
 \label{fig:z7lf}
}
\end{figure}
\begin{figure}
 \centering
 \includegraphics[scale=0.56, viewport=20 0 400 460]{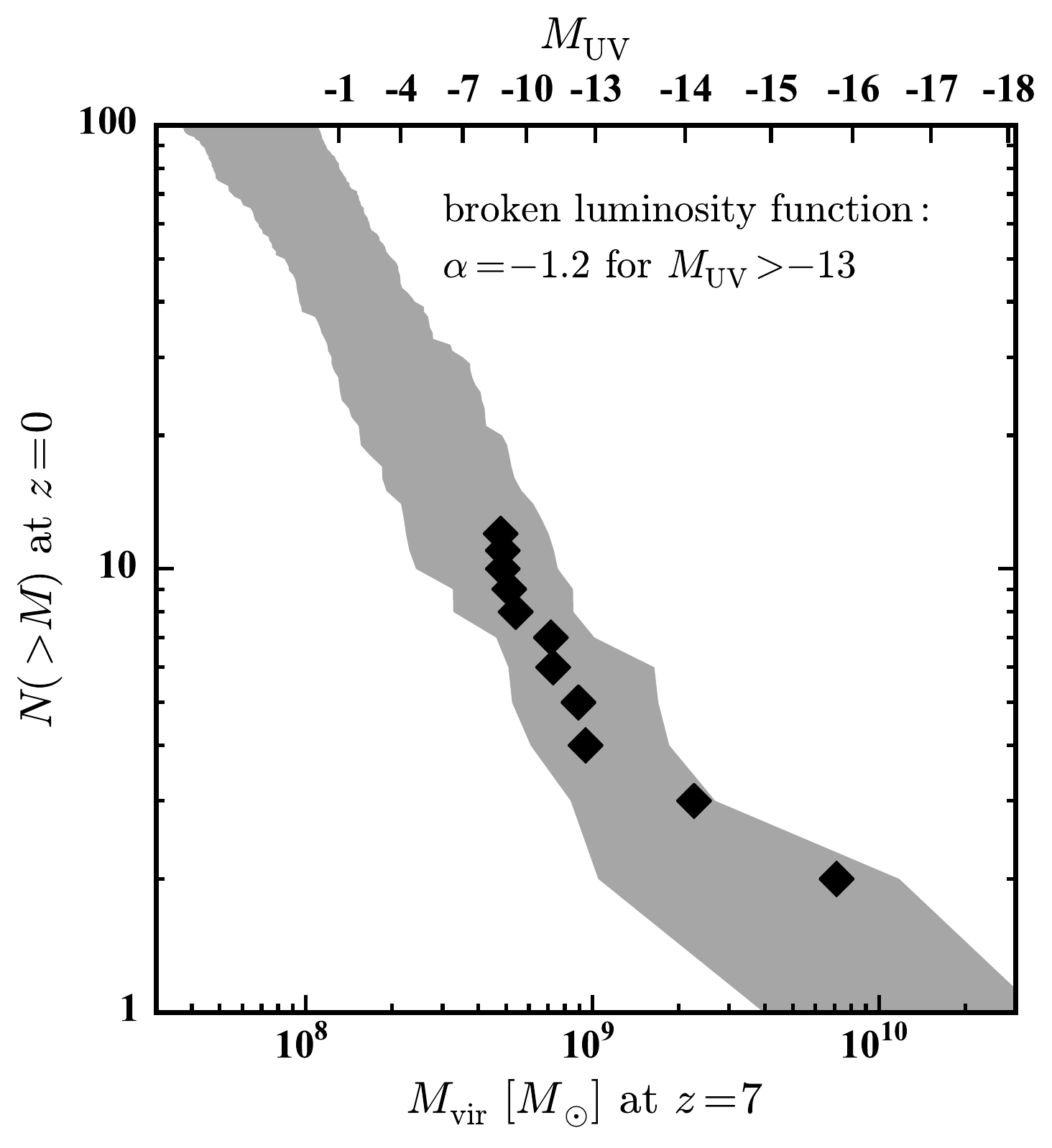}
 \caption{Similar to Figure~\ref{fig:z7lf}, but assumes a UV luminosity function
   that breaks to $\alpha=-1.2$ at $\muv > -13$ (from the fiducial value of
   $\alpha=-2.03$ for brighter galaxies). The $z=7$ census of 
   galaxies surviving to $z=0$ in the Milky Way is in much better agreement with
   the modeled UV 
   luminosities in this case, as galaxies over a wider range in luminosity are
   placed in a narrower range of halo masses (for $\muv > -13$). In such a
   scenario, all known MW galaxies, including ultra-faint dwarfs, could lie at
   or above the atomic cooling threshold of $\mvir \sim 10^8\,\msun$ at $z=7$.
 \label{fig:z7lf_broken}
}
\end{figure}

Figure~\ref{fig:lf_plot_z7} indicates that the faintest galaxies observable in
the HUDF at $z \sim 7$ likely are hosted by
$\mvir \approx 3 \times 10^{10}\,\msun$ halos, while the atomic cooling
threshold of $\tvir \approx 10^4\,{\rm K}$ corresponds to $\muv \approx -10$.
B14 showed that the {\tt ELVIS} suite of simulated Local Groups predicts
approximately 50 surviving, bound remnants of $\mvir(z\sim 7)>10^8\,\msun$ halos
in the Milky Way's virial volume today. They argued this was potentially
problematic, as even low-level star formation in such halos would quickly
over-produce the observed stellar content of Milky Way satellites.

This tension is evident in Figure~\ref{fig:z7lf}, which shows the $z \sim 7$ UV
LF of the Milky Way and its satellites (symbols) as well as predicted dark
matter halo mass functions from the {\tt ELVIS} simulation suite (gray shaded
region). The corresponding values of $\muv$ based on the abundance matching
model described in Sec.~\ref{subsec:z7} are given in the upper horizontal
axis. The LF from direct modeling of SFHs and from abundance matching are in
good agreement for $\muv \la -12$ ($\mvir \approx 5\times 10^{8}\,\msun$), but
the disagreement disappears for fainter galaxies (lower mass halos), with
low-mass halos far outnumbering the number of known galaxies even at the modeled
$z \sim 7$ luminosity of Draco and Leo II ($\muv \sim -9$, corresponding to
$\mhalo \sim 3 \times 10^{7}\,\msun$).  If every dark matter halo is capable of
hosting a galaxy, then there should be 40-100 surviving descendants of galaxies
with $\muv(z\sim7) \la -10$; our modeling predicts there are only 10 or so such
galaxies around the Milky Way today.  Either only a small fraction of the halos
at this mass ($\mhalo \approx 10^8\,\msun$) are capable of cooling gas and
forming stars at $z \sim 7$ or the mapping between halo mass and UV luminosity
is highly stochastic at early times in low-mass halos -- both of which are
contrary to current models and simulation results; or the UV LF breaks at
$\muv \sim -13$, with $\mvir \la 10^9 \,\msun$ halos hosting fainter galaxies
than our fiducial abundance matching model predicts. Whichever of these
possibilities is correct, there are important implications for the threshold of
galaxy formation and the mass scale of halos that host classical and ultra-faint
dSphs.

The most realistic possibility may be that the $z \sim 7$ UV LF breaks at
$\muv \approx -13$, as argued for in B14. Recent simulations have also found LFs
that may flatten at fainter magnitudes than are probed by \hst\
\citep{gnedin2014a, oshea2015}; there is no evidence for the LF flattening in
current \ff\ data (\citealt{atek2015}, Livermore et al. 2015, in preparation),
which reach $\muv \sim -15.5$ at $z \sim 7$. $\muv \approx -13$ is also a
frequently-adopted value of $\mlim$ in models of reionization, although, as
noted above, $\mlim$ need not be associated with any feature or break in the UV
LF.

If a break in the LF is indeed the answer, the Local Group data (in particular,
galaxy counts at $z=0$ combined with modeled UV luminosities at $z \sim 7$)
indicate the faint-end slope for $\muv \ga -13$ should be close to $-1.2$ rather
than the value of approximately $-2$ that is observed in the HUDF. As is shown
in Figure~\ref{fig:z7lf_broken}, fainter galaxies would then live in halos that
are more massive than our original abundance matching prescription would
indicate. This issue is explored in more detail, and at a variety of redshifts,
in Graus et al. (in preparation). In this model (and in
Fig.~\ref{fig:z7lf_broken}), the classical MW dSphs are hosted by halos with
$\mvir(z\sim7)=(0.5-1)\times 10^{9}\,\msun$. All known Milky Way satellites
could therefore be hosted by halos at or above the atomic cooling limit
($\mvir \sim 10^{8}\,\msun$) at $z\sim 7$ (see also
\citealt{milosavljevic2014}).

It is important to emphasize that completeness in the $z=0$ data is not an issue
when constructing Figures~\ref{fig:z7lf}-\ref{fig:z7lf_broken}: we have only
used data for satellites with $L_{\rm V}(z=0) > 10^5\,L_{V,\odot}$ and current
Galactocentric distances of $<300\,\kpc$, a region where our census of
satellites is very likely complete (e.g., \citealt{tollerud2008, koposov2008,
  walsh2009}). Furthermore, the mismatch in Figure~\ref{fig:z7lf} is already
significant for galaxies such as Leo I and Sculptor (with
$L_{V}(z=0) \approx 5 \times 10^{6}\,L_{V,\odot}$). The only galaxy this bright
that has been found within the Milky Way's virial volume since the 1950s is
Sagittarius \citep{ibata1995}, whose presence had been concealed by the Galaxy's
disk. Although we do not have data for Sextans, the discrepancy in numbers shown
in Figure~\ref{fig:z7lf} is an order of magnitude, indicating that even the
inclusion of Sextans and the discovery of several $10^{5}\,L_{V,\odot}$
satellites would not change the qualitative picture described here.

\subsection{Local Group galaxies at cosmic noon}
\begin{figure*}
 \centering
 \includegraphics[scale=0.57]{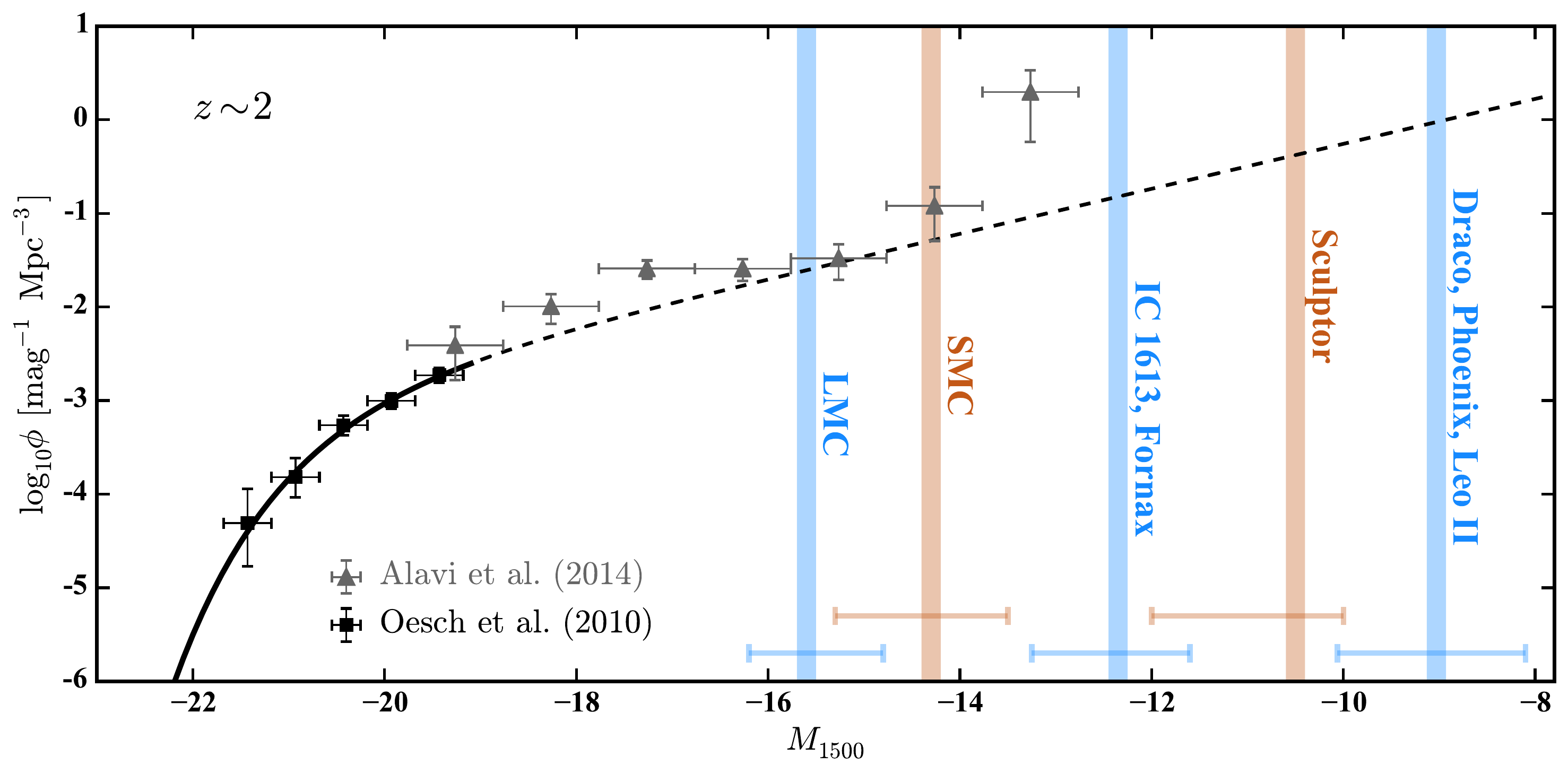}
 \caption{Local Group galaxies in a $z \sim 2$ context. The data points show
   \hst\ observations of $z\sim2$ galaxies in the GOODS-South field
   (\citealt{oesch2010a}; black points) and in the A1689 lensing field
   (\citealt{alavi2014}; gray points). The best-fitting Schechter luminosity
   function from Oesch et al.
   $(M^*=-20.16, \, \phi^*=2.188 \times 10^{-3}\,{\rm mag^{-1}\,Mpc^{-3}}, \,
   \alpha=-1.60)$
   is plotted as a solid curve, and its extrapolation to fainter magnitudes is
   plotted a dashed curve. Vertical bands indicate the modeled
   $\langle \muv(z \approx 2) \rangle$ for various Local Group galaxies (with
   error bars giving $1\,\sigma$ uncertainties; see
   Table~\ref{table:table1}). While deep observations of blank fields only probe
   much more massive galaxies than those of the Local
   Group (except the Galaxy and M31), lensing magnification likely enables the
   study of galaxies similar to progenitors of the LMC and SMC at $z \sim 2$ and
   are nearly capable of reaching the main progenitor of galaxies like IC 1613
   and Fornax at that time.
 \label{fig:lf_plot_z2}
}
\end{figure*}
\label{subsec:z2}
Dwarf galaxies are also important test-beds of galaxy formation physics at eras
other than the epoch of reionization. In particular, large samples of
UV-selected galaxies at $z \sim 2$ are becoming available, and ongoing 
\hst\ programs are enabling studies of galaxies that are intrinsically as faint
as $\muv \sim -13.5$ \citep{alavi2014} at that time, which is close to ``cosmic
noon'', the peak of the cosmic star formation rate density
\citep{madau2014}. Understanding the likely descendants of such galaxies today
-- or the likely progenitors of Local Group galaxies -- will shed further light
on the processes at work in galaxy formation over the past 10 billion years.

Figure~\ref{fig:lf_plot_z2} shows the observed UV LF of galaxies at $z \sim 2$
(black data points, from \citealt{oesch2010a}). The gray data points are
observations that take advantage of lensing magnification combined with deep
near-UV imaging (WFC3/UVIS in F275W), which allowed Alavi et al. to probe to
much fainter galaxies ($\muv \sim -13$) than would otherwise be possible. The
figure also shows the UV magnitudes that a variety of Local Group galaxies would
have at this redshift.

The LMC and SMC are predicted to have $\muv(z\sim2) \approx -16$ and $-14$,
respectively, placing them well within the reach of \hst\ at $z \sim 2$. IC 1613
and the Fornax dSph are predicted to have $\muv(z \sim 2) \approx 12.5$, meaning
the progenitors of these galaxies are just at the edge of \hst's current
capabilities at $z=2$. Other Local Group galaxies in our sample are all
predicted to be fainter than $\muv=-12$ at $z \sim 2$.

As yet larger samples of galaxies become available, it will be possible to
examine $z \sim 2$ galaxies as a population in ever-greater detail. If lensing
allows the study of $\muv \sim -12$ galaxies at $z \sim 2$, a number of
applications will present themselves. Such observations will likely reveal the
progenitors of present-day dwarf irregular galaxies and massive dSphs in a
variety of environments. The distribution of $z\sim2$ UV fluxes obtained through
direct observation will help constrain potential burst parameters (e.g., the
$z \sim 2$ equivalent of Figure~\ref{fig:fornax_burst_muv}). Observations at
$z \sim 2$ will also increase the precision with which the faint end slope of
the UV luminosity function can be measured, placing further constraints on the
evolution of cosmic star formation and on the properties of dust in faint,
star-forming galaxies.

\section{Discussion}
\label{sec:discuss}

The previous sections illustrate the power that deep resolved-star observations
of Local Group galaxies have for understanding not just the nearby Universe but
also (much) earlier eras. Using the modeled UV luminosities, we have shown that
Local Group galaxies such as the Fornax dSph and IC 1613 were, at $z \sim 2$,
slightly less luminous ($\muv \sim -12.5$) than the faintest galaxies \hst\ is
currently observing at that epoch ($\muv \sim -13$). The LMC and SMC are
predicted to be substantially more luminous at that time; many of the faint
($\muv \sim -15$) galaxies observed at $z \sim 2$ may evolve to become
Magellanic irregulars in the local Universe. Pushing to higher redshifts
($z \sim 7$), we find that none of the Local Group galaxies are bright enough to
be seen in the HUDF, consistent with expectations based on number counts and
space densities (e.g., \citealt{trenti2010}, B14).  Many reionization models
predict that extrapolating the observed LF to $\muv \sim -10$ is required to
maintain reionization; if this is the case, the progenitors of some of the
faintest classical dSphs (e.g., Draco, CVn I) and low-luminosity dwarf
irregulars (e.g., Phoenix and Leo A) would be required contributors to the
ionizing background.

In Section~\ref{subsec:near-deep}, we showed that a comparison between our
derived $\muv$ values for Local Group galaxies at $z \sim 7$ and counts of
surviving subhalos in dark-matter-only simulations of the Local Group show that
$z \sim 7$ LF cannot continue with a slope of $\alpha=-2$ to $\muv \sim -9$, as
this would imply an order-of-magnitude excess in surviving satellites with $z=0$
luminosities comparable to the classical MW dwarfs. The faint-end slope derived
from high-$z$ observations is still somewhat degenerate with measured value of
$M^*$ in the Schechter LF, however, with a typical uncertainty of $\pm 0.3$
(e.g., \citealt{mclure2013, finkelstein2014, bouwens2015}). These degeneracies
translate into substantial uncertainties in $\mlim$ (see, e.g., figure 2 of
\citealt{robertson2013}), which in turn will affect the specific value of $\muv$
at which archeological studies in the MW require the high-$z$ LF to break. The
general trend is that if the faint-end slope of the LF is steeper, it must break
at brighter values of $\muv$; if it is shallower, the break can occur at fainter
values. This mirrors the behavior of $\mlim$ and its dependence on $\alpha$.

The Planck collaboration has recently reported a value of the optical depth to
electron scattering, $\tau =0.066 \pm 0.012$ \citep{planck2015}, that is
$\approx 30\%$ lower than previous determinations of $\tau =0.088 \pm 0.014$
\citep{komatsu2011}. \citet{robertson2015} show that extrapolating the LF to a
limiting magnitude of $\mlim \approx -13$ (with a luminosity-independent escape
fraction $\fesc=0.2$) is sufficient to maintain reionization (see also
\citealt{finkelstein2012}) and match the updated Planck determination of
$\tau$. Such a scenario \textit{still} requires extrapolation of the LF down to
progenitors of the Fornax dSph, which has a present-day luminosity of
$L_V\approx 10^7\,L_{V,\odot} $.

If time-averaged escape fractions are lower than 20\% for galaxies with
$\muv \la -13$, as suggested by the work of \citet{wise2014} and \citet{ma2015},
then maintaining reionization would require even fainter galaxies (or additional
sources such as X-rays; e.g., \citealt{mirabel2011}). Galaxy-driven reionization
scenarios therefore still require that most of the known Local Group irregular
galaxies -- and the most luminous of the dSphs -- were necessary contributors to
cosmic reionization. These galaxies are intrinsically very faint at high
redshift: \jwst\ will only resolve galaxies that are an order-of-magnitude
brighter even in a deep field campaign.  The unique high-angular-resolution
capabilities of observatories such as \hst\ and \jwst, and the deep observations
of faint galaxies in and around the Local Group they facilitate, therefore hold
the promise of providing a unique probe of the earliest epoch of galaxy
formation.

\vspace{0.2cm}

\section*{Acknowledgments} 
MB-K thanks Anahita Alavi, Hakim Atek, Steve Finkelstein, and Pascal Oesch for
providing data in tabular form and Brant Robertson, Steve Finkelstein, and Cole
Miller for helpful discussions. Support for this work was provided by NASA
through \hst\ theory grants (programs AR-12836 and AR-13888) from the Space
Telescope Science Institute (STScI), which is operated by the Association of
Universities for Research in Astronomy (AURA), Inc., under NASA contract
NAS5-26555. D.R.W. is supported by NASA through Hubble Fellowship grant
HST-HF-51331.01 awarded by STScI. C.C. is supported by a Packard Foundation
Fellowship. This work used computational resources of the University of Maryland
and those granted the Extreme Science and Engineering Discovery Environment
(XSEDE), which is supported by National Science Foundation grant number
OCI-1053575. Much of the analysis in this paper relied on the python packages
{\tt NumPy} \citep{numpy}, {\tt SciPy} \citep{scipy}, {\tt Matplotlib}
\citep{matplotlib}, and {\tt iPython} \citep{ipython}; we are very grateful to
the developers of these tools. This research has made extensive use of NASA's
Astrophysics Data System and the arXiv eprint service at arxiv.org.

\bibliography{draft}
\label{lastpage}

\end{document}

%% file: extra_preamble.tex
\newcommand{\mvir}{M_{\rm{vir}}}

\newcommand{\tvir}{T_{\rm{vir}}}

\newcommand{\mstar}{M_{\star}}
\newcommand{\msun}{M_{\odot}}

\newcommand{\mpc}{{\rm Mpc}}
\newcommand{\kpc}{{\rm kpc}}

\newcommand{\muv}{{M_{\rm UV}}}
\newcommand{\hst}{\textit{HST}}
\newcommand{\jwst}{\textit{JWST}}
\newcommand{\ff}{Frontier Fields}
\newcommand{\fesc}{f_{\rm esc}}
\newcommand{\mhalo}{M_{\rm halo}}
\newcommand{\mlim}{M_{\rm lim}}

%% file: draft.bbl
\begin{thebibliography}{82}
\expandafter\ifx\csname natexlab\endcsname\relax\def\natexlab#1{#1}\fi

\bibitem[{{Alavi} {et~al.}(2014){Alavi}, {Siana}, {Richard}, {Stark},
  {Scarlata}, {Teplitz}, {Freeman}, {Dominguez}, {Rafelski}, {Robertson}, \&
  {Kewley}}]{alavi2014}
{Alavi}, A. {et~al.} 2014, \apj, 780, 143

\bibitem[{{Alvarez} {et~al.}(2012){Alvarez}, {Finlator}, \&
  {Trenti}}]{alvarez2012}
{Alvarez}, M.~A., {Finlator}, K., \& {Trenti}, M. 2012, \apjl, 759, L38

\bibitem[{{Atek} {et~al.}(2015){Atek}, {Richard}, {Kneib}, {Jauzac},
  {Schaerer}, {Clement}, {Limousin}, {Jullo}, {Natarajan}, {Egami}, \&
  {Ebeling}}]{atek2015}
{Atek}, H. {et~al.} 2015, \apj, 800, 18

\bibitem[{{Ben{\'{\i}}tez-Llambay} {et~al.}(2015){Ben{\'{\i}}tez-Llambay},
  {Navarro}, {Abadi}, {Gottl{\"o}ber}, {Yepes}, {Hoffman}, \&
  {Steinmetz}}]{benitez-llambay2015}
{Ben{\'{\i}}tez-Llambay}, A., {Navarro}, J.~F., {Abadi}, M.~G.,
  {Gottl{\"o}ber}, S., {Yepes}, G., {Hoffman}, Y., \& {Steinmetz}, M. 2015,
  \mnras, 450, 4207

\bibitem[{{Bouwens} {et~al.}(2015{\natexlab{a}}){Bouwens}, {Illingworth},
  {Oesch}, {Caruana}, {Holwerda}, {Smit}, \& {Wilkins}}]{bouwens2015a}
{Bouwens}, R.~J., {Illingworth}, G.~D., {Oesch}, P.~A., {Caruana}, J.,
  {Holwerda}, B., {Smit}, R., \& {Wilkins}, S. 2015{\natexlab{a}},
  {arXiv:1503.08228 [astro-ph]}

\bibitem[{{Bouwens} {et~al.}(2015{\natexlab{b}}){Bouwens}, {Illingworth},
  {Oesch}, {Trenti}, {Labb{\'e}}, {Bradley}, {Carollo}, {van Dokkum},
  {Gonzalez}, {Holwerda}, {Franx}, {Spitler}, {Smit}, \& {Magee}}]{bouwens2015}
{Bouwens}, R.~J. {et~al.} 2015{\natexlab{b}}, \apj, 803, 34

\bibitem[{{Bouwens} {et~al.}(2012){Bouwens}, {Illingworth}, {Oesch}, {Trenti},
  {Labb{\'e}}, {Franx}, {Stiavelli}, {Carollo}, {van Dokkum}, \&
  {Magee}}]{bouwens2012}
---. 2012, \apjl, 752, L5

\bibitem[{{Bovill} \& {Ricotti}(2011)}]{bovill2011}
{Bovill}, M.~S., \& {Ricotti}, M. 2011, \apj, 741, 17

\bibitem[{{Boylan-Kolchin} {et~al.}(2014){Boylan-Kolchin}, {Bullock}, \&
  {Garrison-Kimmel}}]{boylan-kolchin2014}
{Boylan-Kolchin}, M., {Bullock}, J.~S., \& {Garrison-Kimmel}, S. 2014, \mnras,
  443, L44

\bibitem[{{Brown} {et~al.}(2012){Brown}, {Tumlinson}, {Geha}, {Kirby},
  {VandenBerg}, {Mu{\~n}oz}, {Kalirai}, {Simon}, {Avila}, {Guhathakurta},
  {Renzini}, \& {Ferguson}}]{brown2012}
{Brown}, T.~M. {et~al.} 2012, \apjl, 753, L21

\bibitem[{{Bullock} {et~al.}(2000){Bullock}, {Kravtsov}, \&
  {Weinberg}}]{bullock2000}
{Bullock}, J.~S., {Kravtsov}, A.~V., \& {Weinberg}, D.~H. 2000, \apj, 539, 517

\bibitem[{{Cole} {et~al.}(2007){Cole}, {Skillman}, {Tolstoy}, {Gallagher},
  {Aparicio}, {Dolphin}, {Gallart}, {Hidalgo}, {Saha}, {Stetson}, \&
  {Weisz}}]{cole2007}
{Cole}, A.~A. {et~al.} 2007, \apjl, 659, L17

\bibitem[{{Conroy} \& {Gunn}(2010)}]{conroy2010a}
{Conroy}, C., \& {Gunn}, J.~E. 2010, \apj, 712, 833

\bibitem[{{Conroy} {et~al.}(2009){Conroy}, {Gunn}, \& {White}}]{conroy2009a}
{Conroy}, C., {Gunn}, J.~E., \& {White}, M. 2009, \apj, 699, 486

\bibitem[{{Dolphin}(2002)}]{dolphin2002}
{Dolphin}, A.~E. 2002, \mnras, 332, 91

\bibitem[{{Dom{\'{\i}}nguez} {et~al.}(2014){Dom{\'{\i}}nguez}, {Siana},
  {Brooks}, {Christensen}, {Bruzual}, {Stark}, \& {Alavi}}]{dominguez2014}
{Dom{\'{\i}}nguez}, A., {Siana}, B., {Brooks}, A.~M., {Christensen}, C.~R.,
  {Bruzual}, G., {Stark}, D.~P., \& {Alavi}, A. 2014, {arXiv:1408.5788
  [astro-ph]}

\bibitem[{{Duffy} {et~al.}(2014){Duffy}, {Wyithe}, {Mutch}, \&
  {Poole}}]{duffy2014}
{Duffy}, A.~R., {Wyithe}, J.~S.~B., {Mutch}, S.~J., \& {Poole}, G.~B. 2014,
  \mnras, 443, 3435

\bibitem[{{Dunlop} {et~al.}(2013){Dunlop}, {Rogers}, {McLure}, {Ellis},
  {Robertson}, {Koekemoer}, {Dayal}, {Curtis-Lake}, {Wild}, {Charlot},
  {Bowler}, {Schenker}, {Ouchi}, {Ono}, {Cirasuolo}, {Furlanetto}, {Stark},
  {Targett}, \& {Schneider}}]{dunlop2013}
{Dunlop}, J.~S. {et~al.} 2013, \mnras, 432, 3520

\bibitem[{{Finkelstein} {et~al.}(2012){Finkelstein}, {Papovich}, {Ryan},
  {Pawlik}, {Dickinson}, {Ferguson}, {Finlator}, {Koekemoer}, {Giavalisco},
  {Cooray}, {Dunlop}, {Faber}, {Grogin}, {Kocevski}, \&
  {Newman}}]{finkelstein2012}
{Finkelstein}, S.~L. {et~al.} 2012, \apj, 758, 93

\bibitem[{{Finkelstein} {et~al.}(2014){Finkelstein}, {Ryan}, {Papovich},
  {Dickinson}, {Song}, {Somerville}, {Ferguson}, {Salmon}, {Giavalisco},
  {Koekemoer}, {Ashby}, {Behroozi}, {Castellano}, {Dunlop}, {Faber}, {Fazio},
  {Fontana}, {Grogin}, {Hathi}, {Jaacks}, {Kocevski}, {Livermore}, {McLure},
  {Merlin}, {Mobasher}, {Newman}, {Rafelski}, {Tilvi}, \&
  {Willner}}]{finkelstein2014}
---. 2014, {arXiv:1410.5439 [astro-ph]}

\bibitem[{{Fisher} {et~al.}(2014){Fisher}, {Bolatto}, {Herrera-Camus},
  {Draine}, {Donaldson}, {Walter}, {Sandstrom}, {Leroy}, {Cannon}, \&
  {Gordon}}]{fisher2014}
{Fisher}, D.~B. {et~al.} 2014, \nat, 505, 186

\bibitem[{{Freeman} \& {Bland-Hawthorn}(2002)}]{freeman2002}
{Freeman}, K., \& {Bland-Hawthorn}, J. 2002, \araa, 40, 487

\bibitem[{{Garrison-Kimmel} {et~al.}(2014){Garrison-Kimmel}, {Boylan-Kolchin},
  {Bullock}, \& {Lee}}]{garrison-kimmel2014}
{Garrison-Kimmel}, S., {Boylan-Kolchin}, M., {Bullock}, J.~S., \& {Lee}, K.
  2014, \mnras, 438, 2578

\bibitem[{{Girardi} {et~al.}(2002){Girardi}, {Bertelli}, {Bressan}, {Chiosi},
  {Groenewegen}, {Marigo}, {Salasnich}, \& {Weiss}}]{girardi2002}
{Girardi}, L., {Bertelli}, G., {Bressan}, A., {Chiosi}, C., {Groenewegen},
  M.~A.~T., {Marigo}, P., {Salasnich}, B., \& {Weiss}, A. 2002, \aap, 391, 195

\bibitem[{{Girardi} {et~al.}(2010){Girardi}, {Williams}, {Gilbert},
  {Rosenfield}, {Dalcanton}, {Marigo}, {Boyer}, {Dolphin}, {Weisz},
  {Melbourne}, {Olsen}, {Seth}, \& {Skillman}}]{girardi2010}
{Girardi}, L. {et~al.} 2010, \apj, 724, 1030

\bibitem[{{Gnedin} \& {Kaurov}(2014)}]{gnedin2014a}
{Gnedin}, N.~Y., \& {Kaurov}, A.~A. 2014, \apj, 793, 30

\bibitem[{{Governato} {et~al.}(2012){Governato}, {Zolotov}, {Pontzen},
  {Christensen}, {Oh}, {Brooks}, {Quinn}, {Shen}, \& {Wadsley}}]{governato2012}
{Governato}, F. {et~al.} 2012, \mnras, 422, 1231

\bibitem[{{Hidalgo} {et~al.}(2013){Hidalgo}, {Monelli}, {Aparicio}, {Gallart},
  {Skillman}, {Cassisi}, {Bernard}, {Mayer}, {Stetson}, {Cole}, \&
  {Dolphin}}]{hidalgo2013}
{Hidalgo}, S.~L. {et~al.} 2013, \apj, 778, 103

\bibitem[{{Hopkins}(2015)}]{hopkins2015}
{Hopkins}, P.~F. 2015, \mnras, 450, 53

\bibitem[{{Hopkins} {et~al.}(2014){Hopkins}, {Kere{\v s}}, {O{\~n}orbe},
  {Faucher-Gigu{\`e}re}, {Quataert}, {Murray}, \& {Bullock}}]{hopkins2014}
{Hopkins}, P.~F., {Kere{\v s}}, D., {O{\~n}orbe}, J., {Faucher-Gigu{\`e}re},
  C.-A., {Quataert}, E., {Murray}, N., \& {Bullock}, J.~S. 2014, \mnras, 445,
  581

\bibitem[{Hunter(2007)}]{matplotlib}
Hunter, J.~D. 2007, Computing In Science \& Engineering, 9, 90

\bibitem[{{Ibata} {et~al.}(1995){Ibata}, {Gilmore}, \& {Irwin}}]{ibata1995}
{Ibata}, R.~A., {Gilmore}, G., \& {Irwin}, M.~J. 1995, \mnras, 277, 781

\bibitem[{{Johnson} {et~al.}(2013){Johnson}, {Weisz}, {Dalcanton}, {Johnson},
  {Dale}, {Dolphin}, {Gil de Paz}, {Kennicutt}, {Lee}, {Skillman}, {Boquien},
  \& {Williams}}]{johnson2013}
{Johnson}, B.~D. {et~al.} 2013, \apj, 772, 8

\bibitem[{Jones {et~al.}(2001)Jones, Oliphant, Peterson, {et~al.}}]{scipy}
Jones, E., Oliphant, T., Peterson, P., {et~al.} 2001, {SciPy}: Open source
  scientific tools for {Python}, [Online; accessed 2015-02-15]

\bibitem[{{Kauffmann}(2014)}]{kauffmann2014}
{Kauffmann}, G. 2014, \mnras, 441, 2717

\bibitem[{{Komatsu} {et~al.}(2011){Komatsu}, {Smith}, {Dunkley}, {Bennett},
  {Gold}, {Hinshaw}, {Jarosik}, {Larson}, {Nolta}, {Page}, {Spergel},
  {Halpern}, {Hill}, {Kogut}, {Limon}, {Meyer}, {Odegard}, {Tucker}, {Weiland},
  {Wollack}, \& {Wright}}]{komatsu2011}
{Komatsu}, E. {et~al.} 2011, \apjs, 192, 18

\bibitem[{{Koposov} {et~al.}(2008){Koposov}, {Belokurov}, {Evans}, {Hewett},
  {Irwin}, {Gilmore}, {Zucker}, {Rix}, {Fellhauer}, {Bell}, \&
  {Glushkova}}]{koposov2008}
{Koposov}, S. {et~al.} 2008, \apj, 686, 279

\bibitem[{{Kroupa}(2001)}]{kroupa2001}
{Kroupa}, P. 2001, \mnras, 322, 231

\bibitem[{{Kuhlen} \& {Faucher-Gigu{\`e}re}(2012)}]{kuhlen2012a}
{Kuhlen}, M., \& {Faucher-Gigu{\`e}re}, C.-A. 2012, \mnras, 423, 862

\bibitem[{{Ma} {et~al.}(2015){Ma}, {Kasen}, {Hopkins}, {Faucher-Giguere},
  {Quataert}, {Keres}, \& {Murray}}]{ma2015}
{Ma}, X., {Kasen}, D., {Hopkins}, P.~F., {Faucher-Giguere}, C.-A., {Quataert},
  E., {Keres}, D., \& {Murray}, N. 2015, {arXiv:1503.07880 [astro-ph]}

\bibitem[{{Madau} \& {Dickinson}(2014)}]{madau2014}
{Madau}, P., \& {Dickinson}, M. 2014, \araa, 52, 415

\bibitem[{{Madau} {et~al.}(2008){Madau}, {Kuhlen}, {Diemand}, {Moore}, {Zemp},
  {Potter}, \& {Stadel}}]{madau2008}
{Madau}, P., {Kuhlen}, M., {Diemand}, J., {Moore}, B., {Zemp}, M., {Potter},
  D., \& {Stadel}, J. 2008, \apjl, 689, L41

\bibitem[{{Madau} {et~al.}(2014){Madau}, {Weisz}, \& {Conroy}}]{madau2014b}
{Madau}, P., {Weisz}, D.~R., \& {Conroy}, C. 2014, \apjl, 790, L17

\bibitem[{{McConnachie}(2012)}]{mcconnachie2012}
{McConnachie}, A.~W. 2012, \aj, 144, 4

\bibitem[{{McLure} {et~al.}(2013){McLure}, {Dunlop}, {Bowler}, {Curtis-Lake},
  {Schenker}, {Ellis}, {Robertson}, {Koekemoer}, {Rogers}, {Ono}, {Ouchi},
  {Charlot}, {Wild}, {Stark}, {Furlanetto}, {Cirasuolo}, \&
  {Targett}}]{mclure2013}
{McLure}, R.~J. {et~al.} 2013, \mnras, 432, 2696

\bibitem[{{Milosavljevi{\'c}} \& {Bromm}(2014)}]{milosavljevic2014}
{Milosavljevi{\'c}}, M., \& {Bromm}, V. 2014, \mnras, 440, 50

\bibitem[{{Mirabel} {et~al.}(2011){Mirabel}, {Dijkstra}, {Laurent}, {Loeb}, \&
  {Pritchard}}]{mirabel2011}
{Mirabel}, I.~F., {Dijkstra}, M., {Laurent}, P., {Loeb}, A., \& {Pritchard},
  J.~R. 2011, \aap, 528, A149

\bibitem[{{O{\~n}orbe} {et~al.}(2015){O{\~n}orbe}, {Boylan-Kolchin}, {Bullock},
  {Hopkins}, {Ker{\v e}s}, {Faucher-Gigu{\`e}re}, {Quataert}, \&
  {Murray}}]{onorbe2015}
{O{\~n}orbe}, J., {Boylan-Kolchin}, M., {Bullock}, J.~S., {Hopkins}, P.~F.,
  {Ker{\v e}s}, D., {Faucher-Gigu{\`e}re}, C.-A., {Quataert}, E., \& {Murray},
  N. 2015, {arXiv:1502.02036 [astro-ph]}

\bibitem[{{Oesch} {et~al.}(2010){Oesch}, {Bouwens}, {Carollo}, {Illingworth},
  {Magee}, {Trenti}, {Stiavelli}, {Franx}, {Labb{\'e}}, \& {van
  Dokkum}}]{oesch2010a}
{Oesch}, P.~A. {et~al.} 2010, \apjl, 725, L150

\bibitem[{{O'Shea} {et~al.}(2015){O'Shea}, {Wise}, {Xu}, \&
  {Norman}}]{oshea2015}
{O'Shea}, B.~W., {Wise}, J.~H., {Xu}, H., \& {Norman}, M.~L. 2015,
  {arXiv:1503.01110 [astro-ph]}

\bibitem[{P\'erez \& Granger(2007)}]{ipython}
P\'erez, F., \& Granger, B.~E. 2007, Computing in Science and Engineering, 9,
  21

\bibitem[{{Planck Collaboration}(2015)}]{planck2015}
{Planck Collaboration}. 2015, {arXiv:1502.01589 [astro-ph]}

\bibitem[{{Power} {et~al.}(2014){Power}, {Wynn}, {Robotham}, {Lewis}, \&
  {Wilkinson}}]{power2014}
{Power}, C., {Wynn}, G.~A., {Robotham}, A.~S.~G., {Lewis}, G.~F., \&
  {Wilkinson}, M.~I. 2014, {arXiv:1406.7097 [astro-ph]}

\bibitem[{{Ricotti} \& {Gnedin}(2005)}]{ricotti2005}
{Ricotti}, M., \& {Gnedin}, N.~Y. 2005, \apj, 629, 259

\bibitem[{{Ricotti} {et~al.}(2008){Ricotti}, {Gnedin}, \&
  {Shull}}]{ricotti2008}
{Ricotti}, M., {Gnedin}, N.~Y., \& {Shull}, J.~M. 2008, \apj, 685, 21

\bibitem[{{Robertson} {et~al.}(2015){Robertson}, {Ellis}, {Furlanetto}, \&
  {Dunlop}}]{robertson2015}
{Robertson}, B.~E., {Ellis}, R.~S., {Furlanetto}, S.~R., \& {Dunlop}, J.~S.
  2015, \apjl, 802, L19

\bibitem[{{Robertson} {et~al.}(2013){Robertson}, {Furlanetto}, {Schneider},
  {Charlot}, {Ellis}, {Stark}, {McLure}, {Dunlop}, {Koekemoer}, {Schenker},
  {Ouchi}, {Ono}, {Curtis-Lake}, {Rogers}, {Bowler}, \&
  {Cirasuolo}}]{robertson2013}
{Robertson}, B.~E. {et~al.} 2013, \apj, 768, 71

\bibitem[{{Salvaterra} {et~al.}(2011){Salvaterra}, {Ferrara}, \&
  {Dayal}}]{salvaterra2011}
{Salvaterra}, R., {Ferrara}, A., \& {Dayal}, P. 2011, \mnras, 414, 847

\bibitem[{{Schechter}(1976)}]{schechter1976}
{Schechter}, P. 1976, \apj, 203, 297

\bibitem[{{Schenker} {et~al.}(2013){Schenker}, {Robertson}, {Ellis}, {Ono},
  {McLure}, {Dunlop}, {Koekemoer}, {Bowler}, {Ouchi}, {Curtis-Lake}, {Rogers},
  {Schneider}, {Charlot}, {Stark}, {Furlanetto}, \& {Cirasuolo}}]{schenker2013}
{Schenker}, M.~A. {et~al.} 2013, \apj, 768, 196

\bibitem[{{Schmidt} {et~al.}(2014){Schmidt}, {Treu}, {Trenti}, {Bradley},
  {Kelly}, {Oesch}, {Holwerda}, {Shull}, \& {Stiavelli}}]{schmidt2014}
{Schmidt}, K.~B. {et~al.} 2014, \apj, 786, 57

\bibitem[{{Sheth} {et~al.}(2001){Sheth}, {Mo}, \& {Tormen}}]{sheth2001}
{Sheth}, R.~K., {Mo}, H.~J., \& {Tormen}, G. 2001, \mnras, 323, 1

\bibitem[{{Skillman} {et~al.}(2014){Skillman}, {Hidalgo}, {Weisz}, {Monelli},
  {Gallart}, {Aparicio}, {Bernard}, {Boylan-Kolchin}, {Cassisi}, {Cole},
  {Dolphin}, {Ferguson}, {Mayer}, {Navarro}, {Stetson}, \&
  {Tolstoy}}]{skillman2014}
{Skillman}, E.~D. {et~al.} 2014, \apj, 786, 44

\bibitem[{{Stinson} {et~al.}(2007){Stinson}, {Dalcanton}, {Quinn}, {Kaufmann},
  \& {Wadsley}}]{stinson2007}
{Stinson}, G.~S., {Dalcanton}, J.~J., {Quinn}, T., {Kaufmann}, T., \&
  {Wadsley}, J. 2007, \apj, 667, 170

\bibitem[{{Teyssier} {et~al.}(2013){Teyssier}, {Pontzen}, {Dubois}, \&
  {Read}}]{teyssier2013}
{Teyssier}, R., {Pontzen}, A., {Dubois}, Y., \& {Read}, J.~I. 2013, \mnras,
  429, 3068

\bibitem[{{Thomas} {et~al.}(2005){Thomas}, {Maraston}, {Bender}, \& {Mendes de
  Oliveira}}]{thomas2005}
{Thomas}, D., {Maraston}, C., {Bender}, R., \& {Mendes de Oliveira}, C. 2005,
  \apj, 621, 673

\bibitem[{{Tollerud} {et~al.}(2008){Tollerud}, {Bullock}, {Strigari}, \&
  {Willman}}]{tollerud2008}
{Tollerud}, E.~J., {Bullock}, J.~S., {Strigari}, L.~E., \& {Willman}, B. 2008,
  \apj, 688, 277

\bibitem[{{Trenti} {et~al.}(2010){Trenti}, {Stiavelli}, {Bouwens}, {Oesch},
  {Shull}, {Illingworth}, {Bradley}, \& {Carollo}}]{trenti2010}
{Trenti}, M., {Stiavelli}, M., {Bouwens}, R.~J., {Oesch}, P., {Shull}, J.~M.,
  {Illingworth}, G.~D., {Bradley}, L.~D., \& {Carollo}, C.~M. 2010, \apjl, 714,
  L202

\bibitem[{{Van Der Walt} {et~al.}(2011){Van Der Walt}, {Colbert}, \&
  {Varoquaux}}]{numpy}
{Van Der Walt}, S., {Colbert}, S.~C., \& {Varoquaux}, G. 2011, {arXiv:1102.1523
  [astro-ph]}

\bibitem[{{van der Wel} {et~al.}(2011){van der Wel}, {Straughn}, {Rix},
  {Finkelstein}, {Koekemoer}, {Weiner}, {Wuyts}, {Bell}, {Faber}, {Trump},
  {Koo}, {Ferguson}, {Scarlata}, {Hathi}, {Dunlop}, {Newman}, {Dickinson},
  {Jahnke}, {Salmon}, {de Mello}, {Kocevski}, {Lai}, {Grogin}, {Rodney}, {Guo},
  {McGrath}, {Lee}, {Barro}, {Huang}, {Riess}, {Ashby}, \&
  {Willner}}]{van-der-wel2011}
{van der Wel}, A. {et~al.} 2011, \apj, 742, 111

\bibitem[{{Vogelsberger} {et~al.}(2014){Vogelsberger}, {Zavala}, {Simpson}, \&
  {Jenkins}}]{vogelsberger2014b}
{Vogelsberger}, M., {Zavala}, J., {Simpson}, C., \& {Jenkins}, A. 2014, \mnras,
  444, 3684

\bibitem[{{Walsh} {et~al.}(2009){Walsh}, {Willman}, \& {Jerjen}}]{walsh2009}
{Walsh}, S.~M., {Willman}, B., \& {Jerjen}, H. 2009, \aj, 137, 450

\bibitem[{{Watson} {et~al.}(2015){Watson}, {Christensen}, {Knudsen}, {Richard},
  {Gallazzi}, \& {Micha{\l}owski}}]{watson2015}
{Watson}, D., {Christensen}, L., {Knudsen}, K.~K., {Richard}, J., {Gallazzi},
  A., \& {Micha{\l}owski}, M.~J. 2015, \nat, 519, 327

\bibitem[{{Weisz} {et~al.}(2013){Weisz}, {Dolphin}, {Skillman}, {Holtzman},
  {Dalcanton}, {Cole}, \& {Neary}}]{weisz2013}
{Weisz}, D.~R., {Dolphin}, A.~E., {Skillman}, E.~D., {Holtzman}, J.,
  {Dalcanton}, J.~J., {Cole}, A.~A., \& {Neary}, K. 2013, \mnras, 431, 364

\bibitem[{{Weisz} {et~al.}(2014{\natexlab{a}}){Weisz}, {Dolphin}, {Skillman},
  {Holtzman}, {Gilbert}, {Dalcanton}, \& {Williams}}]{weisz2014a}
{Weisz}, D.~R., {Dolphin}, A.~E., {Skillman}, E.~D., {Holtzman}, J., {Gilbert},
  K.~M., {Dalcanton}, J.~J., \& {Williams}, B.~F. 2014{\natexlab{a}}, \apj,
  789, 147

\bibitem[{{Weisz} {et~al.}(2014{\natexlab{b}}){Weisz}, {Johnson}, \&
  {Conroy}}]{weisz2014c}
{Weisz}, D.~R., {Johnson}, B.~D., \& {Conroy}, C. 2014{\natexlab{b}}, \apjl,
  794, L3

\bibitem[{{Weisz} {et~al.}(2012){Weisz}, {Johnson}, {Johnson}, {Skillman},
  {Lee}, {Kennicutt}, {Calzetti}, {van Zee}, {Bothwell}, {Dalcanton}, {Dale},
  \& {Williams}}]{weisz2012a}
{Weisz}, D.~R. {et~al.} 2012, \apj, 744, 44

\bibitem[{{Wheeler} {et~al.}(2015){Wheeler}, {Onorbe}, {Bullock},
  {Boylan-Kolchin}, {Elbert}, {Garrison-Kimmel}, {Hopkins}, \&
  {Keres}}]{wheeler2015}
{Wheeler}, C., {Onorbe}, J., {Bullock}, J.~S., {Boylan-Kolchin}, M., {Elbert},
  O., {Garrison-Kimmel}, S., {Hopkins}, P.~F., \& {Keres}, D. 2015,
  {arXiv:1504.02466 [astro-ph]}

\bibitem[{{Windhorst} {et~al.}(2006){Windhorst}, {Cohen}, {Jansen},
  {Conselice}, \& {Yan}}]{windhorst2006}
{Windhorst}, R.~A., {Cohen}, S.~H., {Jansen}, R.~A., {Conselice}, C., \& {Yan},
  H. 2006, \nar, 50, 113

\bibitem[{{Wise} {et~al.}(2014){Wise}, {Demchenko}, {Halicek}, {Norman},
  {Turk}, {Abel}, \& {Smith}}]{wise2014}
{Wise}, J.~H., {Demchenko}, V.~G., {Halicek}, M.~T., {Norman}, M.~L., {Turk},
  M.~J., {Abel}, T., \& {Smith}, B.~D. 2014, \mnras, 442, 2560

\bibitem[{{Zaritsky} {et~al.}(2012){Zaritsky}, {Colucci}, {Pessev},
  {Bernstein}, \& {Chandar}}]{zaritsky2012}
{Zaritsky}, D., {Colucci}, J.~E., {Pessev}, P.~M., {Bernstein}, R.~A., \&
  {Chandar}, R. 2012, \apj, 761, 93

\bibitem[{{Zolotov} {et~al.}(2012){Zolotov}, {Brooks}, {Willman}, {Governato},
  {Pontzen}, {Christensen}, {Dekel}, {Quinn}, {Shen}, \&
  {Wadsley}}]{zolotov2012}
{Zolotov}, A. {et~al.} 2012, \apj, 761, 71

\end{thebibliography}
